\documentclass[superscriptaddress,twocolumn,pre,amsmath,amssymb, amsfonts]{revtex4-2}
\usepackage{xcolor}
\usepackage{pdfcomment}
\usepackage{comment}
\usepackage{graphicx}
\usepackage{hyperref} 
\hypersetup{
  colorlinks,
  citecolor=blue,
  linkcolor=blue,
  urlcolor=blue
  }

\let\bs\boldsymbol
\DeclareMathOperator{\sgn}{sgn}
\DeclareMathOperator{\smol}{\Omega}
\DeclareMathOperator{\dsmol}{\delta\Omega}
\DeclareMathOperator{\proj}{\mathcal P}
\DeclareMathOperator{\projQ}{\mathcal Q}
\DeclareMathOperator{\Span}{span}
\DeclareMathOperator{\sinc}{sinc}
\DeclareMathOperator{\ce}{ce}
\DeclareMathOperator{\se}{se}

\newcommand{\tlname}[1]{\ensuremath{\text{\textit{#1}}}}

\def\TODO[#1]#2#3{\pdfmarkupcomment[open=true,color=yellow,author={#1}]{#3}{#2}}
\makeatletter
\AtBeginDocument{\immediate\openout\ulp@out=\jobname.upa\relax}
\AtEndDocument{\immediate\write\@auxout{\string\ulp@afterend}}
\makeatother

\usepackage{glossaries}
\makeatletter
\newcommand*{\glsplainhyperlink}[2]{%
  \colorlet{currenttext}{.}
  \colorlet{currentlink}{\@linkcolor}
  \hypersetup{linkcolor=currenttext}
  \hyperlink{#1}{#2}%
  \hypersetup{linkcolor=currentlink}
}
\let\@glslink\glsplainhyperlink
\makeatother
\newacronym{ABP}{ABP}{active Brownian particle}
\newacronym{MCT2}{MCT}{mode-coupling theory}
\newacronym{MCT}{MCT}{mode-coupling theory of the glass transition}
\newacronym{ABPMCT}{ABP-MCT}{mode-coupling theory for active Brownian particles}
\newacronym{ITT}{ITT}{integration-through transients}
\newacronym{MSD}{MSD}{mean-squared displacement}
\newacronym{BD}{BD}{Brownian dynamics}
\newacronym{EDBD}{ED-BD}{event-driven Brownian dynamics}
\newacronym{DFT}{DFT}{density-functional theory}
\newacronym{MHNC}{MHNC}{modified hypernetted-chain}
\newacronym{AOUP}{AOUP}{active Ornstein-Uhlenbeck particles}
\newacronym{SISF}{SISF}{self-intermediate scattering function}

\begin{document}

\title{Mode-Coupling Theory for Tagged-Particle Motion of Active Brownian Particles}
\date\today
\def\dlr{\affiliation{Institut f\"ur Materialphysik im Weltraum,
  Deutsches Zentrum f\"ur Luft- und Raumfahrt (DLR), 51170 K\"oln,
  Germany}}
\def\hhu{\affiliation{Department of Physics,
  Heinrich-Heine Universit\"at D\"usseldorf,
  Universit\"atsstr.~1, 40225 D\"usseldorf, Germany}}
\def\ufr{\altaffiliation[Current address: ]{Physikalisches Institut,
  Albert-Ludwigs-Universit\"at, 79104 Freiburg, Germany}}

\author{Julian Reichert}\dlr
\author{Suvendu Mandal}\hhu\ufr
\author{Thomas Voigtmann}\dlr\hhu

\begin{abstract}
We derive a \acrfull{MCT2}
to describe the dynamics of tracer particles
in dense systems of \glspl{ABP}
in two spatial dimensions. The \gls{ABP} undergo translational and rotational
Brownian dynamics, and are equipped with a fixed self-propulsion speed
along their orientational vector that describes their active motility.
The resulting equations of motion for the tagged-particle density correlation
functions describe the various cases of tracer dynamics close to the glass
transition:
that of a passive
colloidal particle in a suspension of \gls{ABP}, that of a single active particle
in a glass-forming passive host suspensions, and that of active tracers
in a bath of active particles.
Numerical results are presented for these cases assuming hard-sphere
interactions among the particles. The qualitative and quantitative accuracy
of the theory is tested against \acrfull{EDBD}
simulations of active and passive hard disks.
Simulation and theory are found in quantitative agreement,
provided one adjusts the overall density (as known from the
passive description of glassy dynamics), and allows for a rescaling of self-propulsion velocities
in the active host system. These adjustments account for the fact that
\acrshort{ABPMCT} generally overestimates the tendency for kinetic arrest.
We also confirm in the simulations
a peculiar feature of the transient and stationary
dynamical density correlation functions regarding their lack of symmetry
under time reversal, demonstrating the non-equilibrium nature of the system
and how it manifests itself in the theory.
\end{abstract}

\maketitle

\glsresetall

\section{Introduction}

Systems of \glspl{ABP} are paradigmatic models for non-equilibrium
statistical physics \cite{Bechinger.2016}.
They are a conceptually simple toy model that
incorporates both thermalized random Brownian motion, and non-equilibrium
active self-propolsion that captures motion typical of
biological entities such as cells and micro-organisms, as well as of
artificial colloidal microswimmers such as suspended Janus particles driven
by an external energy field \cite{Howse.2007}.

While a lot of attention has been devoted to the dilute and moderately
dense active fluids, the high-density dynamics of \gls{ABP} is still less
explored. Active particles can form states of colective kinetic arrest,
where self-propulsion and steric-hindrance forces balance to form an
``active glass'' \cite{Janssen.2019,Berthier.2019}.
Active glasses have been identified in computer simulations of \gls{ABP}
\cite{Ni.2013}, as well as
in experimental systems employing artificial
microswimmer suspensions that are thought to be good realizations
of the \gls{ABP} model \cite{Klongvessa.2019a,*Klongvessa.2019b,Lozano.2019}.
Also computer models of tissue dynamics show glassy states with similar phenomenology
\cite{Bi.2016},
and in general, collective dynamics and eventual kinetic arrest by glass
formation or jamming of motile entities is considered relevant
for our understanding of mechanisms in the crowded interior of organisms
(see e.g.\ Refs.~\cite{Hoefling.2013,Zhou.2009,Angelini.2011,Parry.2014,Park.2015,Garcia.2015,Oswald.2017,Palamidessi.2019}).

Theoretical studies of active glasses have so far mostly focused on
somewhat simpler model systems, like that of \gls{AOUP}
\cite{Szamel.2015,Szamel.2016,Flenner.2016,Berthier.2018,Feng.2017}
or related models \cite{Berthier.2014,Levis.2015,Mandal.2016,Mandal.2020}.
The activity in the \gls{AOUP} system is modeled through non-white noise with
a finite persistence time, and in
essence this provides for easier theoretical modeling as one does
not have to take into account orientational degrees of freedom
epxlicitly. The model is usually also studied without passive Brownian
motion, i.e., in the athermal limit. Fewer theoretical studies address the dynamics of thermal \gls{ABP} at high densities.
Arguably, to correctly resolve the orientational degrees of freedom
of the \gls{ABP} becomes important at high densities, where the
mean distance between particles is on the order of, or less than the
persistence length of self propulsion.

In the present contribution, we extend our previous approach of using
the \gls{MCT} to describe the high-density dynamics of \gls{ABP}
in two spatial dimensions
\cite{Liluashvili.2017,Liluashvili.2018}.
While the previous work focused on the \emph{collective} dynamics of
density fluctuations near the active-glass transition, here we
complement that theory by equations of motion for the \emph{tagged-particle}
dynamical density correlation functions, to study tracer motion near
the active- and passive-glass transition.
Several points motivate this development: first, the tagged-particle
dynamics is the observable that is most often studied in computer
simulation of glass-forming systems, because of its better statistics.
Second,
the tagged-particle correlation functions provide the theoretical
basis to derive also equations of motion for the \gls{MSD} of tracer
particles, which is the convenient observable to extract from
experimental studies that investigate active suspensions by
direct-imaging techniques \cite{Lozano.2019,GomezSolano.2017}.
Third, many interesting physical effects arise from the interaction
of active with passive particles. This concerns the observation of
active tracers moving in a ``crowded'' environment (previously studied
in models with fixed obstacles \cite{Morin.2016,Morin.2017,Zeitz.2017,Chepizhko.2019,BrunCosmeBruny.2019}), or the motion of active particles in visco-elastic
suspension \cite{GomezSolano.2016,Narinder.2018,Narinder.2019}.
Also, the embedding of passive probe particles in active fluids provides
an interesting case in light of experiments, specifically micro-rheology
techniques that assess the dynamics of biological fluids by observation
of suspended colloidal particles \cite{Wirtz.2009,Bursac.2005,Nishizawa.2017}.
Such passive tracers have been observed to undergo super-diffusive motion
as a clear sign of the non-equilibrium forces active in the bath
\cite{Wu.2000,Caspi.2000,Chen.2007,Wilhelm.2008,Gal.2010,Valeriani.2011,Lagarde.2020}.
The effective interactions of passive particles provided by an active bath
have also been studied recently in experiment \cite{Semeraro.2018}.

A further central topic of the present paper is to assess the
accuracy of the \gls{ABPMCT} in comparison to computer simulation. To this
end we employ \gls{EDBD} computer simulations \cite{Scala.2007,Ni.2013}
as a tool to implement the Brownian dynamics of particles with strict hard-sphere
no-overlap interactions, and compare the \gls{SISF} in various setups of
active and passive tracer particles in active and passive host systems.
This provides a crucial test of the approximations invoked in the theory,
and establishes the validity of \gls{ABPMCT} for a regime of dense
suspensions whose slow structural relaxation is modified by activity.
The simulations also allow to test peculiar features of the theoretically
predicted correlation functions that relate to the inherent
non-equilibrium nature of the model system.

The paper is structured as follows: in Sec.~\ref{sec:theory} we recapitulate
the \gls{ABPMCT} for the collective dynamics, and extend it to include
equations of motion for the tagged-particle dynamics of a tracer with possibly
different interactions and different activity parameters than the host
system. In Sec.~\ref{sec:results}, we demonstrate numerical solutions
of the equations of motion for the tracer-correlation functions, and
compare the \gls{ABPMCT} results to those of our \gls{EDBD} simulations:
after establishing the baseline in a comparison of passive-in-passive
tracer dynamics (Sec.~\ref{sec:results:passive}), we turn to the motion of an active tracer first in the dilute system (Sec.~\ref{sec:results:dilute}), then in the passive host system (Sec.~\ref{sec:results:active_in_passive}), and finally we discuss
the motion of passive and active tracers in the dense active host system (Secs.~\ref{sec:results:passive_in_active} and \ref{sec:results:active_in_active}).
A small note on the differences between transient and stationary averages
in the correlation functions (Sec.~\ref{sec:results:transient}) follows.
Section~\ref{sec:conclusion} provides concluding remarks.

\section{Theory}\label{sec:theory}

\subsection{Mode-Coupling Theory}

The \gls{ABPMCT} describes tagged-particle motion in \gls{ABP} as coupled
to the dynamics of collective density fluctuations; the corresponding equations
for the latter have been derived by Liluashvili et~al.\ \cite{Liluashvili.2017}
and shall be repeated here for completeness.

We study systems of $N$ \gls{ABP} in two spatial dimensions, with
positions $\vec r_k$ and orientation angles $\varphi_k$ ($k=1,\ldots N$).
The stochastic equations of motion are
\begin{subequations}\label{eq:abp}
\begin{align}
  d\vec r_k&=\mu\vec F_k\,dt+\sqrt{2D_t}\,d\vec W_k+v_0\vec n(\varphi_k)\,dt\,,\\
  d\varphi_k&=\sqrt{2D_r}\,dW_{\varphi_k}\,,
\end{align}
\end{subequations}
where the $\vec F_k$ are the direct interaction forces.
The Brownian motion is driven by independent Wiener processes $d\vec W_k$
and $dW_{\varphi_k}$, where $D_t$ and $D_r$ are the respective diffusion
coefficients. The particle mobility $\mu$ is chosen to obey the
equilibrium detailed-balance condition, $\mu=\beta D_t$, where
$\beta=1/kT$ is the inverse temperature.
Active motion is modeled as a fixed self-propulsion of
velocity $v_0$ along the orientation vector of each particle,
$\vec n(\varphi_k)=\vec n_k=(\cos\varphi_k,\sin\varphi_k)^T$.
This term is formally treated as an (arbitrarily strong) non-equilibrium
perturbation to the passive equilibrium dynamics.

The theory will be applied to particles that experience spherically symmetric
direct interactions akin to those of hard spheres.
We fix the units of length and time via the particle diameter $\sigma$ and
the translational diffusion time $\sigma^2/D_t$. This leaves two fundamental
parameters to characterize the \gls{ABP} motion, viz., the rotational
diffusion coefficient $D_r$ and the self-propulsion velocity $v_0$.

Translating the equations of motion, Eqs.~\eqref{eq:abp},
according to the theory of Markov processes one obtains the (backward)
Smoluchowski operator $\smol^\dagger$ that drives the time evolution
of observables,
\begin{equation}\label{eq:smol}
  \smol^\dagger=\sum_{k=1}^ND_t(\vec\nabla_k+\beta\vec F_k)\cdot\vec\nabla_k
  +D_r\partial_{\varphi_k}^2+v_0\vec n_k\cdot\vec\nabla_k\,.
\end{equation}
The Smoluchowski operator is the sum of an equilibrium operator and
an active driving term, $\smol^\dagger=\smol^\dagger_\text{eq}+\dsmol^\dagger$,
with $\dsmol^\dagger=v_0\sum_k\vec n_k\cdot\vec\nabla_k$.
Where convenient, we split off the rotational part of the operator,
$\smol^\dagger_R=D_r\sum_k\partial_{\varphi_k}^2$, denoting the remainder
as its translational part, $\smol^\dagger_T=\smol^\dagger-\smol^\dagger_R$.
The direct interaction forces are assumed to derive from a potential,
$\vec F_k=-\vec\nabla_kU$, so that $p_\text{eq}\propto\exp[-\beta U]$
is the passive-equilibrium solution of the Smoluchowski equation for
$v_0=0$.

The central quantity of \gls{ABPMCT} are the angle-resolved density fluctuations,
$\varrho_l(\vec q)=\sum_{k=1}^N\exp[i\vec q\cdot\vec r_k]\exp[il\varphi_k]/\sqrt{N}$,
and the corresponding transient dynamical correlation functions
\begin{equation}\label{eq:phi}
\Phi_{ll'}(\vec q,t)=\langle\varrho_l^*(\vec q)\exp[\smol^\dagger t]
\varrho_{l'}(\vec q)\rangle_\text{eq}\,.
\end{equation}
Here, angular brackets $\langle\cdot\rangle$
denote the usual \emph{equilibrium} ensemble average, formed with the
Boltzmann weight $p_\text{eq}$,
while the time evolution is understood to contain the
full non-equilibrium dynamics.
The subscript ``eq'' is implicit in the following, except in cases where
we compare equilibrium and non-equilibrium averages.
The transient correlation functions are at the heart of the \gls{ITT}
formalism to derive expressions for non-equilibrium transport coefficients
that are of generalized Green-Kubo type. In principle, they are accessible
in experiment and simulation when the non-equilibrium perturbation --
the particles' activity in our case -- is suddenly switched on starting
from a passive-equilibrium state.

A Mori-Zwanzig projection operator scheme
allows to derive an exact equation of motion of the transient density
correlation function,
by projecting the dynamics onto the angle-resolved
density fluctuations. One gets \cite{Liluashvili.2017}, in matrix notation with respect to the
angular-mode indices,
\begin{multline}\label{eq:mzphi}
  \partial_t\bs\Phi(\vec q,t)+\bs\omega(\vec q)\cdot\bs S^{-1}(q)
  \cdot\bs\Phi(\vec q,t)
  \\
  +\int_0^tdt'\,\bs m(\vec q,t-t')\cdot\left(\bs1\partial_{t'}
  +\bs\omega_R\right)\cdot\bs\Phi(\vec q,t')=\bs0\,,
\end{multline}
to be solved with initial condition $\bs\Phi(\vec q,0)=\bs S(q)$. Here
$S_{ll'}(q)=\delta_{ll'}S_l(q)$ is the equilibrium static structure factor
of the passive system, assumed to be a known input quantity to the theory.
Since we assume the interaction potential between the particles to be
spherically symmetric, the only non-trivial term is $S_0(q)\equiv S(q)$, while
$S_l(q)=1$ for all $l\neq0$.
In general, we will assume the system to remain
homogeneous and isotropic, so that the positional ($l=l'=0$) correlation
functions depend on the wave vector $\vec q$ only through its magnitude
$q=|\vec q|$.

The matrix $\bs\omega(\vec q)=\bs\omega_T(\vec q)+\bs\omega_R
=-\langle\varrho_l^*(\vec q)\smol^\dagger\varrho_{l'}(\vec q)\rangle$
is the (negative) matrix element of the Smoluchowski operator,
separated into its
contributions to the translational and the rotational degrees of freedom,
\begin{subequations}\label{eq:omega}
\begin{align}
  \omega_{T,ll'}(\vec q)&=\delta_{ll'}q^2D_t-\delta_{|l-l'|,1}
  \frac{iqv_0}{2}e^{-i(l-l')\theta_q}S_l(q)\,,\\
  \omega_{R,ll'}&=\delta_{ll'}l^2D_r\,.
\end{align}
\end{subequations}
Here, $\vec q=q(\cos\theta_q,\sin\theta_q)^T$. The non-trivial properties
of the solutions of Eq.~\eqref{eq:mzphi} for active particles (as compared
to the established solutions for passive particles) stem from
the fact that $\bs\omega_T(\vec q)$ is a tri-diagonal matrix and
couples different $l$-modes.

The memory kernel $\bs m(\vec q,t)=\bs M(\vec q,t)\cdot\bs\omega_T^{-1}(\vec q)$
is the focus of \gls{MCT} approximations. Assuming that the major contribution
to these retarded friction effects on the motion of particles stems from
the overlap of fluctuating forces with density-pair modes, \gls{MCT}
sets
\begin{multline}
  M_{l_1l_2}(\vec q,t)\approx\frac{n}{2}\frac1{(2\pi)^2}
  \int d\vec k\,\sum_{l_3l_4l_{3'}l_{4'}}
  \mathcal V^\dagger_{l_1l_3l_4}(\vec q,\vec k,\vec p)
  \times\\ \times
  \Phi_{l_3l_{3'}}(\vec k,t)
  \Phi_{l_4l_{4'}}(\vec p,t)\mathcal V_{l_2l_{3'}l_{4'}}(\vec q,\vec k,\vec p)
\end{multline}
with $n=N/V$ the density of the system, $\vec q=\vec k+\vec p$, and vertices
$\mathcal V^\dagger=\mathcal V+\delta\mathcal V^\dagger$. The vertex
$\mathcal V$ is the same as in the passive-equilibrium theory,
\begin{subequations}
\begin{multline}\label{eq:Veq}
  \mathcal V_{l_1l_3l_4}(\vec q,\vec k,\vec p)
  =D_t\delta_{l_1,l_3+l_4}\Bigl[(\vec q\cdot\vec k)c_{l_3}(k)\\
  +(\vec q\cdot\vec p)c_{l_4}(p)\Bigr]\,,
\end{multline}
while the active contribution enters only $\mathcal V^\dagger$,
\begin{multline}
  \delta\mathcal V^\dagger_{l_1l_3l_4}(\vec q,\vec k,\vec p)
  =
  \frac{iv_0}{2}\delta_{|l_1-l_3-l_4|,1}S_{l_1}(q)
  \times\\ \times
  \Bigl[ke^{-i(l_1-l_3-l_4)\theta_k}S_{l_1-l_4}(k)\left(
  c_{l_1-l_4}(k)-c_{l_3}(k)\right)\\
  +pe^{-i(l_1-l_3-l_4)\theta_p}S_{l_1-l_3}(p)\left(
  c_{l_1-l_3}(p)-c_{l_4}(p)\right)\Bigr]\,.
\end{multline}
\end{subequations}
In these expressions, $c_l(q)$ is the direct correlation function,
related to the static structure factor by
$S_l(q)=(1-nc_l(q))^{-1}$. In particular, $c_l(q)=0$ for all $l\neq0$.

\subsection{Mode-Coupling Theory for Tagged-Particle Motion}

We now consider a tagged (tracer)
particle embedded in the $N$-particle system described above, with
position $\vec r_s$ and orientation $\varphi_s$. Its equation of motion
shall be as Eq.~\eqref{eq:abp}, where the three relevant parameters
are the diffusion coefficients $D_t^s$ and $D_r^s$, and the
tagged-particle self-propulsion velocity $v_0^s$.
The interaction forces between the tracer and the host-system particles
can in principle also differ from the one among the host particles; for
example, the tracer could be of different size.
For simplicity, we focus here on tracers of equal size and
interactions as the host particles, $\sigma^s=\sigma$, and all parameters
that are not explicitly given are taken as identical to those of the host
system. Note however that we
specifically allow for the case of a passive tracer in an active bath,
or vice versa, i.e., of specific interest is the case $v_0^s\neq v_0$.

The tagged-particle
density fluctuations $\varrho^s_l(\vec q)=\exp[i\vec q\cdot\vec r_s]
\exp[il\varphi_s]$ define the tagged-particle transient density correlation
function
\begin{equation}\label{eq:phis}
  \phi^s_{ll'}(\vec q,t)=\langle\varrho^{s*}_l(\vec q)
  \exp[\smol^\dagger t]\varrho^s_{l'}(\vec q)\rangle\,,
\end{equation}
where $\smol^\dagger$ is given by Eq.~\eqref{eq:smol} with the sum
over particles extended to include the tagged particle.
The positional density-correlation function $\phi^s_{00}(q,t)$ is
also referred to as the \acrfull{SISF}.
The derivation of the \gls{MCT} equations of motion for $\bs\phi^s(\vec q,t)$
proceeds in analogy to the derivation for the collective counter-part
$\bs\Phi(\vec q,t)$:

A Mori-Zwanzig projection onto the tagged-particle density fluctuations,
using the projection operator $\proj=\sum_l\varrho^s(\vec q)\rangle
\langle\varrho_l^{s*}(\vec q)$ together with the Dyson decomposition
\begin{multline}\label{eq:projs}
  \partial_t\exp[\smol^\dagger t]
  =\smol^\dagger\proj\exp[\smol^\dagger t]
  +\smol^\dagger\projQ\exp[\smol^\dagger\projQ t]\\
  +\smol^\dagger\projQ\int_0^tdt'\,\exp[\smol^\dagger\projQ(t-t')]
  \smol^\dagger\proj\exp[\smol^\dagger t']\,,
\end{multline}
yields an equation of motion for the density-correlation function: inserting
Eq.~\eqref{eq:projs} into Eq.~\eqref{eq:phis},
\begin{multline}\label{eq:mzs1}
  \partial_t\bs\phi^s(\vec q,t)=-\bs\omega^s(\vec q)\cdot\phi^s(\vec q,t)\\
  +\int_0^tdt'\,\bs K^s(\vec q,t-t')\cdot\bs\phi^s(\vec q,t')
\end{multline}
where $\bs\omega^s(\vec q)$ is given by the analog of Eq.~\eqref{eq:omega},
\begin{subequations}\label{eq:omegas}
\begin{gather}\label{eq:omegasT}
  \omega_{T,ll'}^s(\vec q)=\delta_{ll'}q^2D_t^s
  -\delta_{|l-l'|,1}\frac{iqv_0^s}{2}e^{-i(l-l')\theta_q}\,,\\
  \omega_{ll'}^s(\vec q)=\omega_{T,ll'}^s(\vec q)+l^2D_r^2\delta_{ll'}\,.
\end{gather}
\end{subequations}
The Mori-Zwanzig memory kernel
is given by
\begin{equation}\label{eq:Ks}
  K^s_{ll'}(\vec q,t)=\left\langle\varrho_l^s(\vec q)^*
  \smol^\dagger\projQ\exp[\smol^\dagger\projQ t]\smol^\dagger
  \varrho_{l'}^s(\vec q)\right\rangle\,.
\end{equation}
In this expression, the spherical symmetry of the direct particle
interactions imposes $\smol^\dagger_R\varrho^s_{l'}(\vec q)
\in\Span_m\{\varrho_m^s(\vec q)\}$, and together with the fact that
$\smol^\dagger_R$ is self-adjoint with respect to the scalar product
defined by the equilibrium average, it implies that
to both sides of the exponential in Eq.~\eqref{eq:Ks}, one can replace
$\smol^\dagger$ by $\smol^\dagger_T$.

Equation~\eqref{eq:mzs1} is treated further in order to rewrite the
memory kernel $\bs K^s(\vec q,t)$ into a friction memory kernel:
we set $\proj'=-\sum_{ll'}\varrho^s_l(\vec q)\rangle
(\bs\omega_T^s)^{-1}_{ll'}\langle\varrho^{s*}_{l'}(\vec q)\smol^\dagger_T$
to take out the one-particle reducible dynamics by a further Dyson
decomposition,
\begin{multline}
  \exp[\smol^\dagger\projQ t]=\exp[\smol^\dagger\projQ'\projQ t]
 \\
  +\int_0^tdt'\,\exp[\smol^\dagger\projQ(t-t')]\smol^\dagger_T
  \proj'\projQ\exp[\smol^\dagger\projQ'\projQ t']\,.
\end{multline}
Inserting into Eq.~\eqref{eq:Ks} results in
\begin{equation}\label{eq:KsMs}
  \bs K^s(t)=\bs M^s(t)
  -\int_0^tdt'\,\bs K^s(t-t')\cdot\bs\omega_T^{s,-1}
  \cdot\bs M(t')
\end{equation}
(dropping the $\vec q$ dependence for notational convenience).
Equations \eqref{eq:mzs1} and \eqref{eq:KsMs} can be combined to a single
equation of motion for the density correlator,
\begin{multline}\label{eq:mcts}
  \partial_t\bs\phi^s(\vec q,t)+\bs\omega^s(\vec q)\cdot\bs\phi^s(\vec q,t)
  \\
  +\int_0^tdt'\,\bs m^s(\vec q,t-t')\cdot
  \left(\bs1\partial_{t'}+\bs\omega^s_R\right)\cdot\bs\phi^s(\vec q,t')
  =\bs 0\,,
\end{multline}
where we have set $\bs m^s(t)=\bs M^s(t)\cdot\bs\omega_T^{s,-1}$.
The exact microscopic expression for the irreducible memory kernel is
as in Eq.~\eqref{eq:Ks}, with the replacement of the projected time-evolution
operator by the further reduced one,
\begin{equation}\label{eq:Ms}
  M_{ll'}^s(\vec q,t)
  =\left\langle\varrho_l^s(\vec q)\smol_T^\dagger\projQ
  \exp[\projQ\smol^\dagger\projQ'\projQ t]\projQ\smol_T^\dagger
  \varrho_{l'}^s(\vec q)\right\rangle\,.
\end{equation}

\Gls{MCT} proceeds by approximating this memory kernel in terms of
density pair modes. For the description of tagged-particle motion, the
first non-trivial overlap is with the mixed modes
$\varrho_{l_3}(\vec k)\varrho^s_{l_4}(\vec p)$, so that inserting a
projector onto such pairs to both sides of the exponential operator
in Eq.~\eqref{eq:Ms}, and splitting the resulting dynamical four-point
correlation function into a product of two-point correlation functions,
one obtains
\begin{multline}\label{eq:Msmct}
  M^s_{l_1l_2}(\vec q,t)\approx\frac{n}{(2\pi)^2}\int d\vec k
  \sum_{l_3l_4}\mathcal W_{l_1l_3l_4}^s(\vec q,\vec k)
  \times\\ \times
  \Phi_{l_3,0}(\vec k,t)\phi^s_{l_4l_2}(\vec p,t)\,.
\end{multline}
In this expression, we have anticipated that the
equilibrium part of the vertex is obtained in analogy to
Eq.~\eqref{eq:Veq}, but due to the assumption of a tagged particle
that is separate from the $N$ host particles, a further symmetry under
$\vec k\leftrightarrow\vec p$ allows
to reduce the integral to twice that over the first term in the
vertex expression. Making further use of the symmetry of the interaction
potential, which ensures that all $c_l(q)$ with $l\neq0$ vanish,
the tagged-particle vertex can be summarized as
\begin{multline}\label{eq:Ws}
  \mathcal W_{l_1l_3l_4}^s(\vec q,\vec k)
  =(D_t^s c^s(k))^2\Bigl[\delta_{l_1l_4}\delta_{l_30}(\vec q\cdot\vec k)^2
  \\
  +\delta_{|l_1-l_3-l_4|,1}\frac{i(\vec q\cdot\vec k)ke^{-i(l_1-l_3-l_4)\theta_k}}
  {2D_t^s}(v_0S(k)\delta_{l_1l_4}-v_0^s\delta_{l_30})
  \Bigr]\,.
\end{multline}
The tagged-particle direct correlation function $c^s(k)$ quantifies the
interactions between the tracer particle and the host-system particles.
In the simplest case of a structually identical (but possibly different
in activity) tracer, $c^s(k)=c(k)$.

Note that in Eq.~\eqref{eq:Ws} \emph{both} the tagged-particle self-propulsion
velocity $v_0^s$, and the host system's $v_0$ enter. Thus the dynamics of a
passive tracer in an active host system is modified both indirectly, through
an implicit $v_0$-dependence of the correlation function $\Phi_{l,0}(\vec k,t)$
appearing in Eq.~\eqref{eq:Msmct}, and directly through a modified
coupling vertex.

In summary, the tagged-particle correlation functions are determined
from a \gls{MCT} memory equation with a memory kernel
$\bs M^s(\vec q,t)$ that couples bi-linearly to the collective
density correlators $\bs\Phi(\vec k,t)$, and the tagged-particle
correlators $\bs\phi^s(\vec p,t)$. Thus, Eqs.~\eqref{eq:mcts}--\eqref{eq:Ws} can be solved
once the collective dynamics has been determined.

\subsection{Small-Wavenumber Limit}

The \gls{SISF} is connected to the
\gls{MSD} of the tracer particle in the low-$q$ limit,
\begin{equation}\label{eq:phismsd}
  \phi^s_{00}(q,t)=1-\frac{q^2}4\delta r^2(t)+\mathcal O(q^4)\,,
\end{equation}
where the factor $4$ holds in two spatial dimensions. It is thus instructive to discuss the
$q\to0$ limit of the \gls{MCT} memory kernel $\bs m^s(\vec q,t)$. We
include its derivation here for completeness, although a detailled
discussion of the \gls{MSD} and its comparison to experiment
is outside the scope of the present contribution
and is relegated to separate publications \cite{msdpaper,Reichert.2021}.

The derivation is simplified by noting the transformation rule for the
correlation functions under a rotation of $\vec q$. Invariance of the
equilibrium distribution and the Smoluchowski operator under a rotation
of the coordinate system implies $\bs\Phi(\vec q^{\,\prime},t)=\bs u^\dagger(\psi)
\cdot\bs\Phi(\vec q,t)\cdot\bs u(\psi)$ if $\vec q^{\,\prime}$ is obtained
from rotating $\vec q$ by an angle $\psi$ and $u_{ll'}(\psi)=\delta_{ll'}\exp[il\psi]$ is the unitary representation of the rotation group. The
quantities
\begin{equation}
  \tilde\Phi_{ll'}(q,t)=e^{i(l-l')\theta_q}\Phi_{ll'}(\vec q,t)
\end{equation}
are thus functions of the wave vector through $q=|\vec q|$ only.
We define corresponding isotropized matrices for the other quantities
that appear in the equations of motion and denote them by tildes.
It is readily checked that the Mori-Zwanzig equations of motion are
invariant under this mapping.
There further holds $\Phi^*_{-l,-l'}(\vec q,t)=(-)^{l-l'}\Phi_{ll'}(\vec q,t)$,
and it follows that the diagonal elements are isotropic and even functions
of $q$. In particular, $\Phi_{00}(q,t)$ is indeed isotropic and real-valued,
in agreement with Eq.~\eqref{eq:phismsd}.

Since $\tilde{\bs m}^s(q,t)=\tilde{\bs M}^s(q,t)\cdot\tilde{\bs\omega}_T^{s,-1}(q)$,
the discussion of the low-$q$ limit requires an expression for the inverse of
$\tilde{\bs\omega}_T^s(q)$. Note that the latter is a symmetric
tri-diagonal matrix,
\begin{equation}\label{eq:omegaTs}
  \tilde\omega_{T,ll'}^s(q)=\delta_{ll'}D_t^sq^2-\delta_{|l-l'|,1}\frac{iqv_0^s}2\,.
\end{equation}
This simple structure allows to derive a closed form for the matrix elements
of its inverse,
\begin{equation}\label{eq:omegaTsinv}
  \left(\tilde{\bs\omega}_T^s\right)^{-1}_{ll'}
  =\frac1\Delta
  \left(\frac{iqv_0^s}{D_t^sq^2+\Delta}\right)^{|l-l'|}
\end{equation}
setting $\Delta=\sqrt{(D_t^sq^2)^2+(v_0^sq)^2}$.
This is readily checked by direct computation of $\tilde{\bs\omega}_T^s\cdot
\tilde{\bs\omega}_T^{s,-1}$. A direct derivation of this result is
given in Appendix~\ref{sec:omegainv}.

It is worth stressing that implicit in this result is the recognition that
the Mori-Zwanzig equations based on the angle-resolved density fluctuations
are formulated for elements $\phi^s_{ll'}(\vec q,t)$ of an infinite-dimensional
matrix algebra spanned by the angular indices. It is customary (and
necessary for numerical computation) to introduce some angular-index
cutoff $L$ such that $l,l'\in[-L,L]$. However, the operations of introducing
such cutoff and of taking the inverse in the limit $q\to0$ do not commute
for $\tilde{\bs\omega}_T^s(q)$. Numerical evaluation in fact confirms that
for $v_0^s\neq0$, the inversion of the
finite-dimensional matrix yields an asymptotic behavior $\sim1/q^2$ for
the $(00)$ element of the inverse for all even cutoff values $L$,
while the same element
approaches a constant for all $L$ odd. On the infinite-dimensional
algebra, Eq.~\eqref{eq:omegaTsinv} confirms the asymptote $1/(v_0^sq)$
for $l=l'=0$ and $v_0^s\neq0$. Obtaining this asymptote is crucial in
deriving the correct $q\to0$ limit of the \gls{ABPMCT} equations.

It is also remarkable that the $q\to0$ limit of Eq.~\eqref{eq:omegaTsinv}
does not commute with $v_0^s\to0$. In physical terms, any small tracer
activity is eventually felt at large enough length scales, i.e., if
$q\ll q_*=v_0^s/D_t^s$. Expansion of Eq.~\eqref{eq:omegaTsinv} yields
\begin{equation}\label{eq:omegaTsinvq0}
  \left(\tilde{\bs\omega}_T^s\right)^{-1}_{ll'}
  \sim\begin{cases}
  1/(D_t^sq^2)\delta_{ll'} & \text{for $v_0^s=0$,}\\
  \frac1{v_0^sq}i^{|l-l'|}\left(1-|l-l'|\frac{q}{q_*}\right)+\mathcal O(q)
  & \text{for $v_0^s\neq0$.}
  \end{cases}
\end{equation}

It requires some care to check that the $q\to0$ expansion of the memory
integral in Eq.~\eqref{eq:mcts} remains regular and produces terms
of leading order $\mathcal O(q^2)$ in the case of the $(00)$ element,
as required by Eq.~\eqref{eq:phismsd}. In the equilibrium theory,
this is readily noted by realizing that the vertex $\mathcal W^s$,
Eq.~\eqref{eq:Ws} is of $\mathcal O(q^2)$, so that
$\tilde m^s_{00}(q,t)\sim\mathcal O(q^0)$ combines with the
$\mathcal O(q^2)$ order of $\partial_{t'}\phi^s_{00}(q,t)$. The appearance
of non-diagonal terms in both $\tilde{\bs m}^s(q,t)$ and $\tilde{\bs\phi}^s(q,t)$, and of a potentially ``dangerous'' term of $\mathcal O(q)$ in
the vertex $\mathcal W^s$,
complicate matters in the \gls{ABPMCT}. After checking the different
contributions individually, as detailed in Appendix~\ref{sec:lowqdetail},
one obtains an equation for the \gls{MSD},
\begin{multline}\label{eq:msd}
  \partial_t\delta r^2(t)+\int_0^tdt'\,\hat m^s_{00}(t-t')\partial_{t'}\delta r^2(t')\\
  =4D_t^s-2iv_0^s\sum\nolimits_\pm\hat\phi^s_{\pm1,0}(t)\\
  +4\sum\nolimits_\pm\int_0^tdt'\,\hat m^s_{0,\pm1}(t-t')
  (\partial_{t'}+D_r^s)\hat\phi^s_{\pm1,0}(t')\,,
\end{multline}
where $\hat m_{ll'}(t)=\lim_{q\to0}\tilde m_{ll'}(q,t)/q^{|l-l'|}$
for $|l-l'|\le1$.
Equation~\eqref{eq:msd} reduces to the
\gls{MCT} equation for the \gls{MSD} of a passive tracer
\cite{Fuchs.1998,Bayer.2007} when $v_0^s=0$,
where the non-diagonal elements of the correlators vanish.
In the case of an active tracer particle, a further equation determines
the off-diagonal low-$q$ correlator, $\hat\phi^s_{\pm1,0}(t)=\lim_{q\to0}\tilde\phi^s_{\pm1,0}(q,t)/q$,
\begin{multline}\label{eq:msd1}
  (\partial_t+D_r)\hat\phi^s_{\pm1,0}(t)=\frac{iv_0^s}2\\
  -2\int_0^tdt'\,\hat m^s_{\pm1,\pm1}(t-t')(\partial_{t'}+D_r)\hat\phi^s_{\pm1,0}(t')\,.
\end{multline}
Explicit expressions for the memory kernels appearing in Eqs.~\eqref{eq:msd} and \eqref{eq:msd1} are given in Appendix~\ref{sec:lowqdetail}.

Equations \eqref{eq:msd} and \eqref{eq:msd1} determine the \gls{MSD}
of an active or passive tracer in an active of passive bath, once the
corresponding tagged-particle correlation functions at finite $\vec q$
have been determined and with these the low-$q$ memory kernels $\hat{\bs m}(t)$.
As we discuss elsewhere \cite{msdpaper,Reichert.2021}, in the non-interacting
case where $\hat m_{ll'}^s(t)\equiv0$, Eqs.~\eqref{eq:msd} and \eqref{eq:msd1}
confirm the well-known analytical expression for the \gls{MSD} of a
free \gls{ABP}.

\subsection{Numerical Solution of MCT Equations}

The \gls{ABPMCT} equations were solved numerically with an algorithm
outlined in Ref.~\cite{Liluashvili.2017}, using a discrete wave-number
grid $q_i=(i+1)\Delta q/2$, $i=1,\ldots M$ with a grid size of $M=128$
and step size $\Delta q\simeq0.3$, implying a large-wave-number cutoff
$Q=40 \sigma$. The integrals are performed for $\vec q$ placed along
the $y$-axis, and the correlation functions for different wave-vector
orientations that are required in the memory-kernel integrals are obtained
by their known transformation behavior under spatial rotations
\cite{Liluashvili.2017}.
Angular-mode indices were considered with a cutoff
$l\in[-L,L]$ with $L=1$, which is deemed sufficient for the range
of self-propulsion velocities that we study within the theory here.

We apply the theory to hard-disk systems, where the only relevant state
parameter in equilibrium is the packing fraction, $\varphi=(\pi/4)n\sigma^2$.
The equilibrium static structure factor of the 2D hard-disk system
was obtained from \gls{DFT}
\cite{Thorneywork.2018}.
This differs slightly from the structure factor used in previous studies
of two-dimensional hard disks in \gls{MCT}, where either \gls{MHNC}
has been used \cite{Bayer.2007} or the empirical Baus-Colot formula
\cite{Liluashvili.2017}.
No qualitative changes are imposed by these different choices; but the
predicted numerical value of the
glass-transition density depends both on the choice of $S(k)$
and the parameters of the discretization in the wave-vector integrals.

\subsection{Brownian Dynamics Simulations}

To compare the theoretical predictions with simulation results, we
performed \gls{EDBD} simulations of size-polydisperse hard-disk systems
with $N=1000$ particles (adjusting the box size to achieve the desired
packing fraction for any given realization of polydispersity).
Particles sizes were drawn randomly from a uniform distribution of
width $0.1\sigma$.

The \gls{EDBD} algorithm has been described in detail for passive systems
\cite{Scala.2007}. For every Brownian time step of length $\tau_B$,
the particles are assigned random Gaussian displacements $\Delta x$
with a variance that sets the short-time
diffusion coefficient. To resolve particle overlaps, the displacements are
translated into pseudo-velocities, $v=\Delta x/\tau_B$, and an event-driven
molecular-dynamics simulation segment of length $\tau_B$ is performed
with elastic collisions between the particles. This algorithm reproduces
the Smoluchowski dynamics of the passive system if $\tau_B$ is chosen
not too large \cite{Scala.2007}.

The algorithm is extended ad~hoc to active particles by assigning Gaussian
displacements to the particles with an individual drift corresponding to
the active driving term. For this, also particle orientation angles are
tracked and incremented by Gaussian displacements to establish rotational
diffusion. A version of this algorithm has first been used by Ni et~al.\
\cite{Ni.2013} for studying the high-density dynamics of \gls{ABP}.

For each density we performed at least 200 runs to accumulate statistics for the
dynamical correlation functions. The systems were equilibrated first, and
then the self-propulsion term was switched on and the systems were brought
into stationary state by preparation runs of at least $10^4$ Brownian time steps.
To study the tracer dynamics, an individual particle was selected within
each simulated configuration to closely match the average size of the
particles (in order to avoid size-disparity effects that become visible
when averaging the motion of tracers of different sizes in the different
simulation runs). To improve the statistics, correlation functions were
averaged over different starting points in the stationary simulations.

We hence compare in the following the \emph{stationary} correlation
functions obtained from the simulation with the \emph{transient} ones
obtained by \gls{ABPMCT}, under the salient asumption that the two
quantities are at least qualitatively similar.
For the fully active system, the simulation statistics was also sufficient
to calculate the transient dynamical correlation functions, obtained from
the simulation-ensemble average over the passive-equilibrium starting
configurations. We will for this case discuss the similarities and
differences between the two quantities.

\section{Results}\label{sec:results}

\subsection{Passive Dynamics}\label{sec:results:passive}

\begin{figure}
\includegraphics[width=\linewidth]{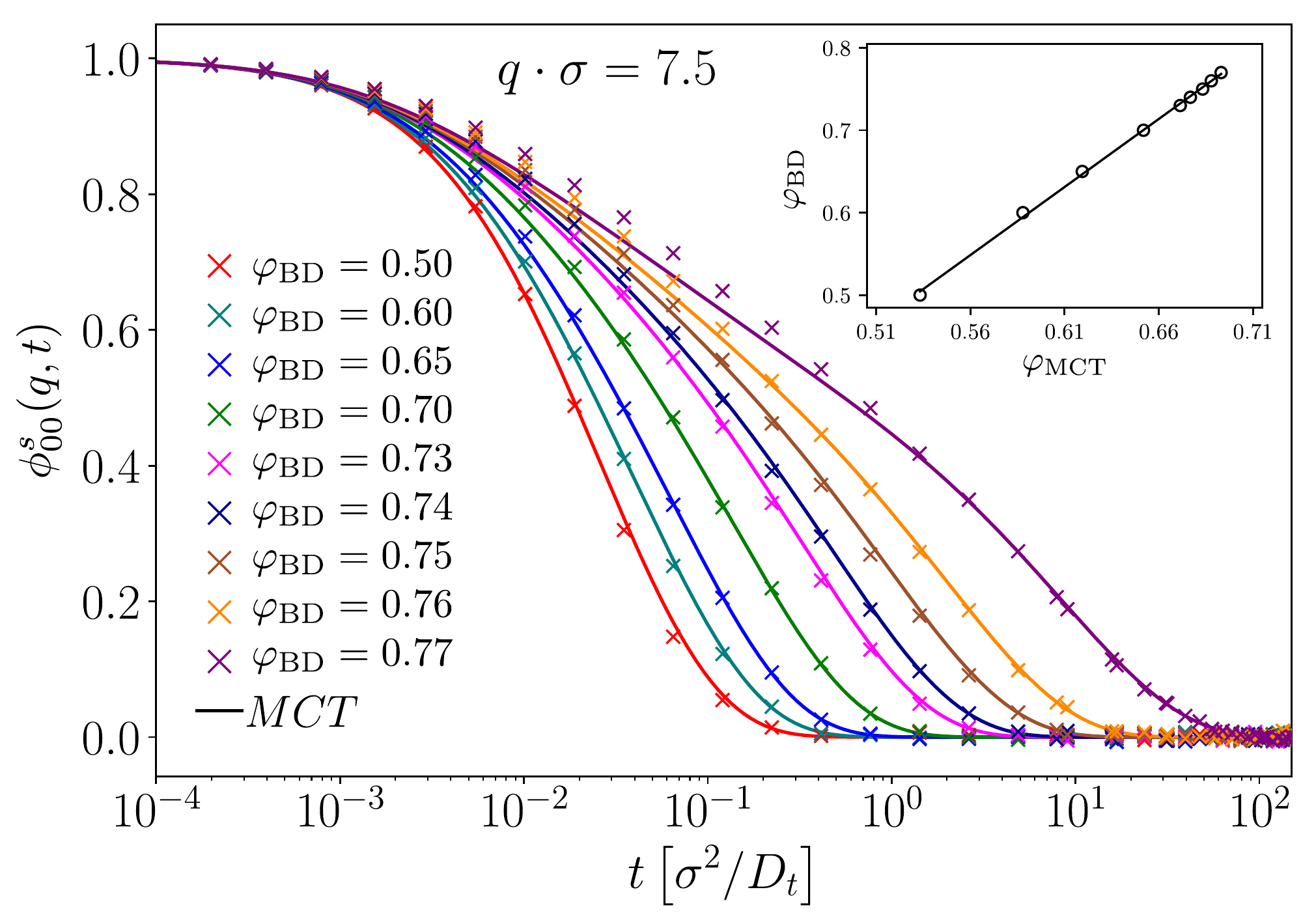}
\caption{\label{fig:passivefit}
  Tagged-particle density correleation function $\phi^s_{00}(q,t)$ for
  passive hard-disk systems of packing fraction $\varphi_\text{BD}$
  as indicated, at wave number $q\sigma=7.5$. Symbols
  are results from Brownian dynamics computer simulations, lines are
  MCT results for packing fractions $\varphi_\text{MCT}$ adjusted to give the
  best description of the data (see inset for the linear relation between
  $\varphi_\text{BD}$ and $\varphi_\text{MCT}$.
}
\end{figure}

We begin by establishing the level of accuracy that can be expected from
\gls{MCT} by comparing to our simulations of a passive hard-disk system.
It is well known that \gls{MCT} provides a semi-quantitative description
of the relaxation dynamics in a density range just below the glass-transition
density, after an adjustment of packing fractions \cite{Weysser.2010}
that accounts for errors both in $S(q)$ and those inherent to \gls{MCT}.

With such an adjustment of the packing fraction,
our solutions of the MCT equations for passive hard disks indeed match the computer
simulation results very well. Anticipating that the glassy dynamics is
driven by fluctuations around the nearest-neighbor peak in the static
structure factor, the adjustment of packing fractions is performed by
treating $\varphi_\text{MCT}$ as a free fit parameter such that
at wave number $q\sigma=7.5$ the long-time structural relaxation matches best
that of the simulation at a given packing fraction $\varphi_\text{BD}$.
Due to its better sampling statistics, we chose to fit the
tagged-particle density correlation function rather than its collective
counterpart; at this wave number, no essential change is expected from
this \cite{Weysser.2010}.
In agreement with earlier studies of similar spirit \cite{Weysser.2010,Bayer.2007} 
the thus adjusted MCT gives a quantitative account of the dynamics
(Fig.~\ref{fig:passivefit}).
Moreover, a free fit of the obtained $\varphi_\text{MCT}$-vs-$\varphi_\text{BD}$
relation for a range of simulations
at $\varphi_\text{BD}\ge0.5$, confirms a linear relationship
between the two quantities (inset of Fig.~\ref{fig:passivefit}).
Considering that within MCT asymptotically close to the glass transition,
a distance parameter that is asymptotically linearly related to the
control-parameter distance governs the long-time dynamics, the linear
relationship found from the free fit corroborates that the adjustment
procedure simply accounts for a well-understood error of MCT in predicting
quantitatively the numerical value of the glass-transition point.

\begin{figure}
\includegraphics[width=\linewidth]{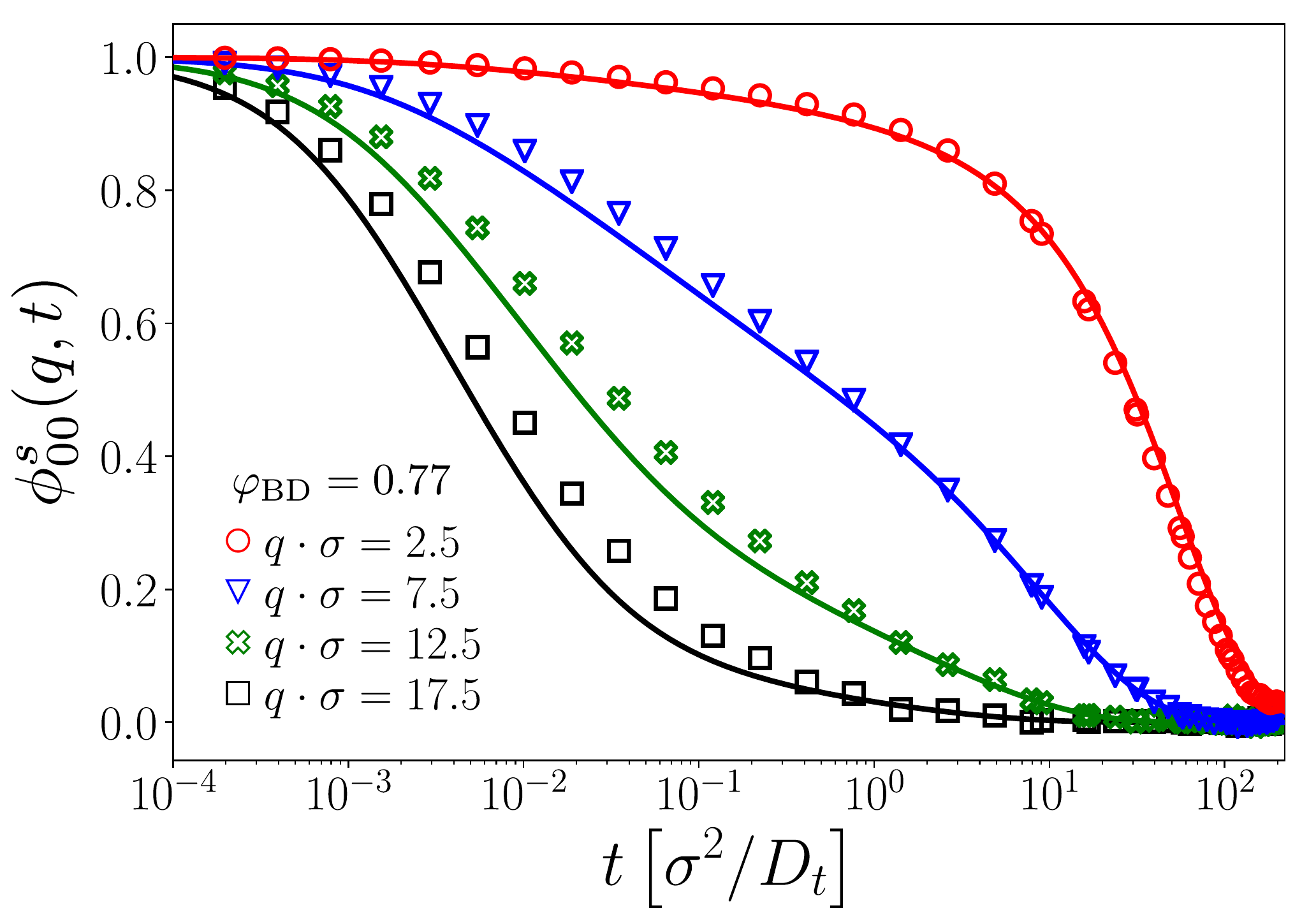}
\caption{\label{fig:passivefitotherq}
  Tagged-particle density corrleation function $\phi^s_{00}(q,t)$ for
  passive hard-disk systems of packing fraction $\varphi_\text{BD}=0.77$,
  for different wave numbers as
  indicated. Symbols are BD simulation results, lines are MCT with effective
  packing fractions. 
}
\end{figure}

Once the mapping of packing fractions is established for the density-fluctuation
dynamics of the relevant length scale, MCT also accounts well for the
dynamics at other wave numbers in the range covering the relevant variations
in $S(q)$ (Fig.~\ref{fig:passivefitotherq}). The agreement remains very good
down to $q\sigma\approx2.5$ in our simulations, although it is known from
similar comparisons in three-dimensional hard-sphere-like systems that for
still smaller $q$, deviations systematically set in \cite{Weysser.2010}.
Also as found previously in 3D, the description of the short-time dynamics
deteriorates at large $q$. These effects appear similarly in our 2D
comparison and establish the relevant $q$ range over which MCT is expected
to quantitatively account for the dynamics.

\subsection{Low-Density Active Dynamics}\label{sec:results:dilute}

\begin{figure}
\includegraphics[width=\linewidth]{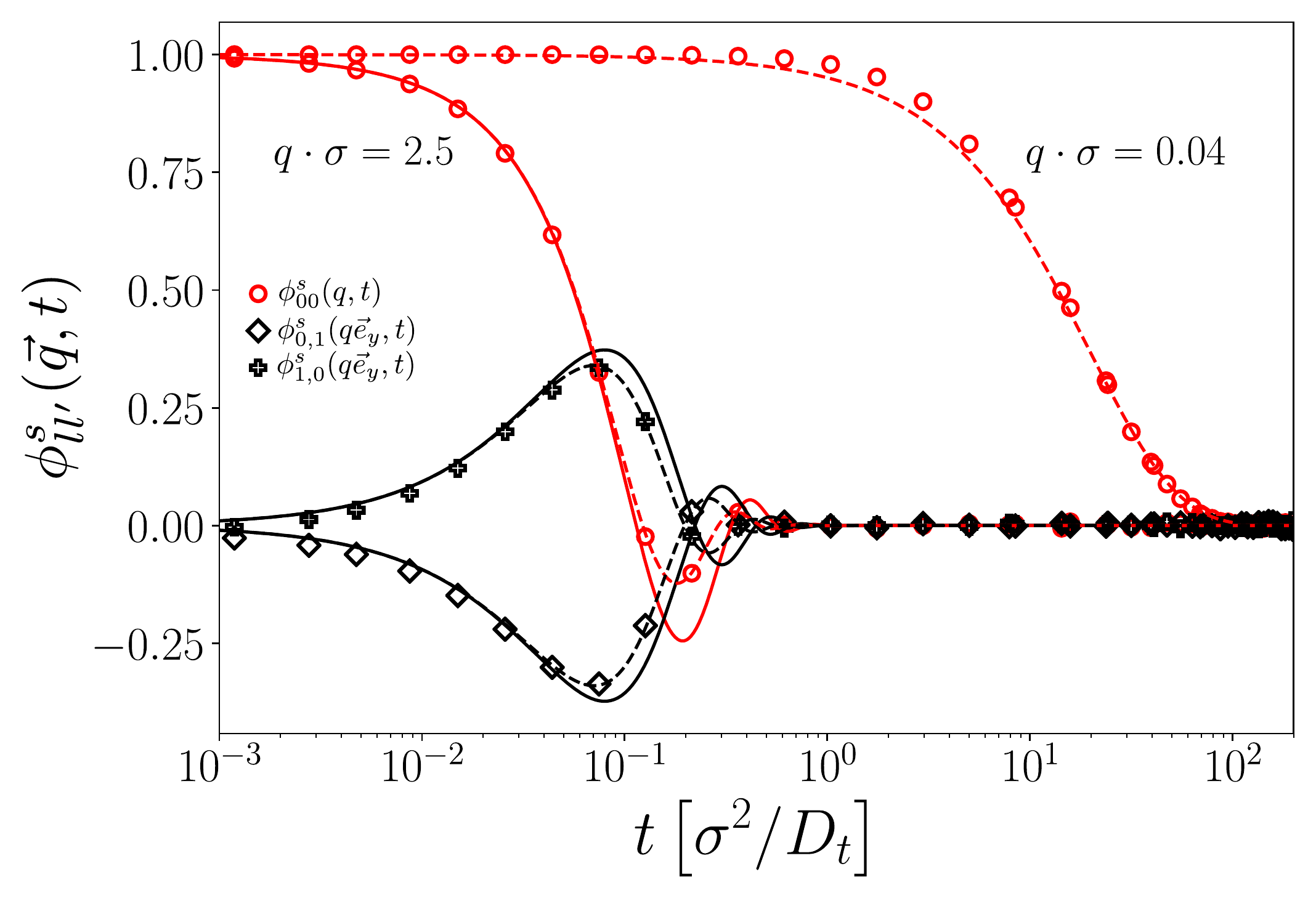}
\caption{\label{fig:lowdens}
  Correlation functions $\phi^s_{ll'}(\vec q,t)$ for a single active Brownian
  particle with self-propulsion velocity $v_0^s=8\,D_t/\sigma$ and
  rotational diffusion coefficient $D^s_r=1\,D_t/\sigma^2$, at a wave vector
  $\vec q\cdot\sigma=2.5\vec e_y$, for $(ll')=(00)$ (circles),
  $(1,0)$ (crosses), and $(0,1)$ (diamonds).
  Symbols are BD simulation results;
  solid lines are the solution of the Mori-Zwanzig equations with
  angular-mode cutoff $L=1$, dashed lines with $L=10$.
  For $\vec q\cdot\sigma=0.04\vec e_y$, $\phi^s_{00}(q,t)$ is shown from
  the simulation (circles) together with the diffusive asymptote
  (dashed line; see text).
}
\end{figure}

The Mori-Zwanzig equation is an exact rewriting of the equations of motion,
and with our choice of projection onto all $\{\varrho_m(\vec q)\}_{m\in\mathbb Z}$,
the memory kernel is at least of $\mathcal O(n)$. (The same does not hold
if one projects only on the positional density fluctuations $\varrho_0(\vec q)$.)
The low-density solution is therefore given by dropping the memory kernel
in Eq.~\eqref{eq:mcts} and thus as a matrix exponential,
\begin{equation}\label{eq:phimatrixexp}
  \bs\phi^s(\vec q,t)=\exp[-\bs\omega^s(\vec q)t]\,.
\end{equation}
The matrix exponential has to be evaluated numerically with some cutoff
on the infinite-dimensional matrices, and exemplary solutions for different
cutoff $L$ are compared with simulation results in Fig.~\ref{fig:lowdens}.

The features of the $(ll')=(00)$ correlation functions of a single \gls{ABP}
have been discussed in detail by Kurzthaler et~al.\ \cite{Kurzthaler.2016,Kurzthaler.2018}.
While passive Brownian motion leads to a decay $\sim\exp(-q^2D^s_tt)$, the
persistence length of self propulsion sets an intermediate-$q$ regime, where
the correlation function decays non-monotonically $\approx\sinc(qv_0^st)$
(exemplified by the $q\sigma=2.5$ curve in Fig.~\ref{fig:lowdens}).
For $q\to0$, the active motion leads to effective enhanced diffusion,
$\sim\exp(-q^2D^s_{t,\text{eff}}t)$ with $D^s_{t,\text{eff}}=D^s_t(1+\tlname{Pe})$
and the P\'eclet number $\tlname{Pe}=(v_0^s)^2/2D_t^sD_r^2$ ($q\sigma=0.04$
in Fig.~\ref{fig:lowdens}).

Equation~\eqref{eq:phimatrixexp} readily provides also the solutions
for $(ll')\neq(00)$.
For example, the functions
for $(ll')=(10)$ and $(01)$ display a clear maximum around $t=2/(qv_0^s)$
and are, for $\vec q$ chosen along $\vec e_y$, positive and negative
mirror images of each other (crosses and diamonds in Fig.~\ref{fig:lowdens}),
as can be anticipated by inspecting Eq.~\eqref{eq:omegas}.

The analytical solution for the ($ll')=(00)$ \gls{SISF} by
Kurzthaler et~al.\ \cite{Kurzthaler.2018}
proceeds by an expansion of the characteristic function of the stochastic
process in terms of suitable eigenfunctions of the single-particle
Smoluchowski operator, the Mathieu functions (in 2D).
The extension to $(ll')\neq(00)$ in this representation
is straightforward, and we present the
calculation in Appendix~\ref{sec:mathieu}.
The two analytical approaches necessarily need to coincide;
although the equivalence is somewhat tedious to show
\cite{ReichertPhD}. We confirm in Appendix~\ref{sec:mathieu} that the
Mathieu function basis indeed diagonalizes the matrix $\bs\omega^s(\vec q)$.
It is nevertheless worth checking the quality of
the Mori-Zwanzig solution, Eq.~\eqref{eq:phimatrixexp}, with a finite $l$-cutoff.
For the range of parameters that are of interest here, indeed
already a cutoff $L=1$ gives good qualitative agreement (solid lines in
Fig.~\ref{fig:lowdens}), although larger $L$ would be needed to
increase the quantitative description
of the oscillatory decay of the correlation function in a $q$-range
that probes the length scales of persistent motion of the \gls{ABP}.
Convergence is slow in this regime: a cutoff of $L=10$ (dashed lines
in Fig.~\ref{fig:lowdens}) would be required to approximate the exact
solutions to within line width in the figure. Our simulation results then
coincide quantitatively with the matrix-exponential solution.

\subsection{Active Tracer in a Passive Suspension}\label{sec:results:active_in_passive}

\begin{figure}
\includegraphics[width=\linewidth]{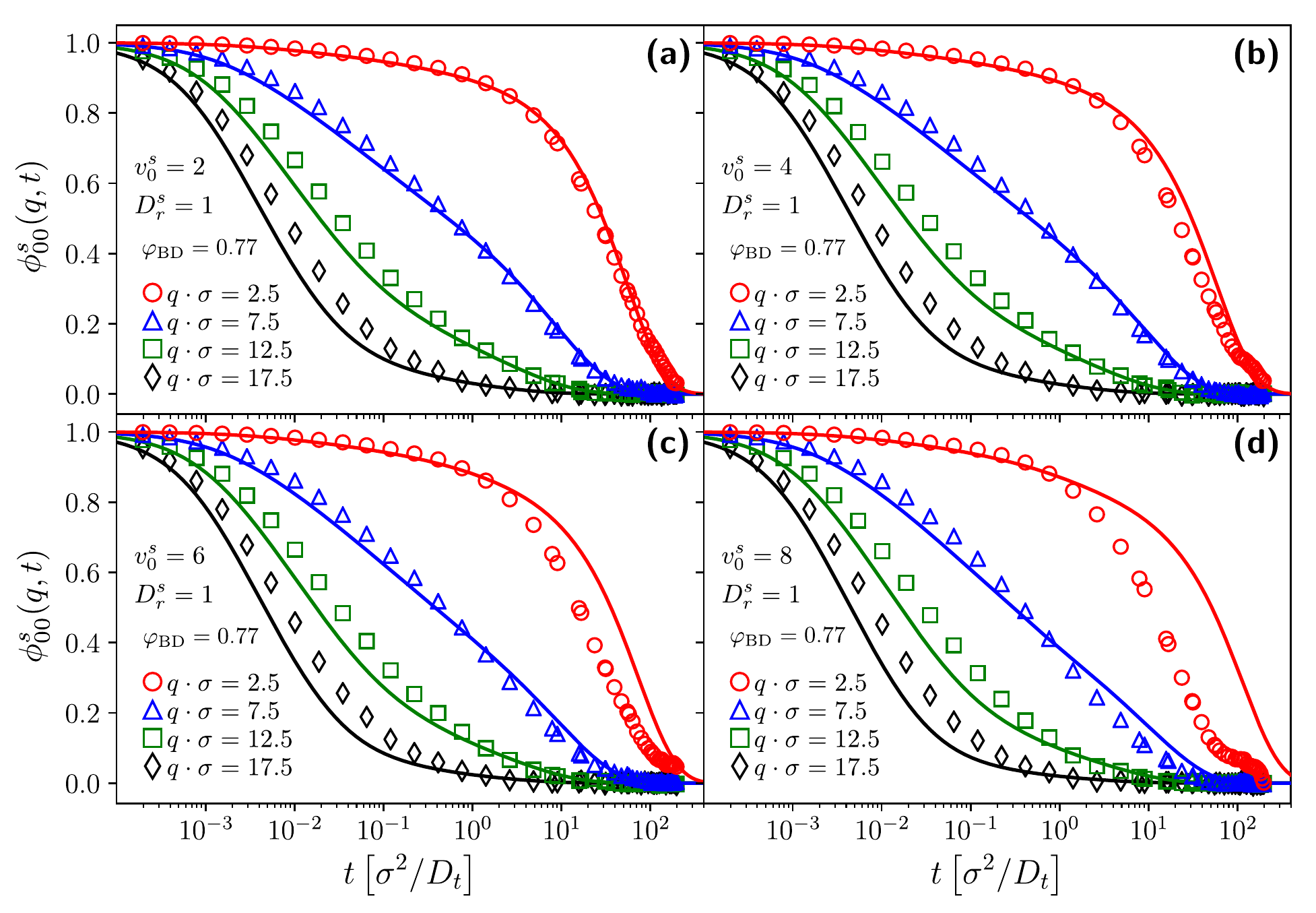}
\caption{\label{fig:active-passive}
  Self-intermediate scattering function of an active tracer in a passive bath, BD simulations (open symbols)
  compared with MCT (lines), at different wave numbers as indicated, at
  fixed packing fration $\varphi=0.77$.
  (a)--(d) exemplify different self-propulsion velocities of the tracer,
  $v_0^s=2$, $4$, $6$, and $8\,D_t/\sigma$.
}
\end{figure}

A first test of the inclusion of self-propulsion forces in \gls{ABPMCT}
is provided by the dynamics of a single ABP embedded in a
dense host suspension of passive particles. The tracer dynamics
decays faster with increasing self-propulsion velocity, $v_0^s$, of the
tracer, although the overall influence of self propulsion in the dense
host suspension is relatively weak (Fig.~\ref{fig:active-passive}).
\gls{ABPMCT} somewhat underestimates this effect, but in particular at
wave numbers $q$ related to the structure-factor maximum, the theory is
in good agreement with the simulation. Deviations become most pronounced
for low $q$ and large $v_0^s$; the deviations at low $q$ are in fact
expected already on the basis of passive \gls{MCT} \cite{Weysser.2010}.

\begin{figure}
\includegraphics[width=\linewidth]{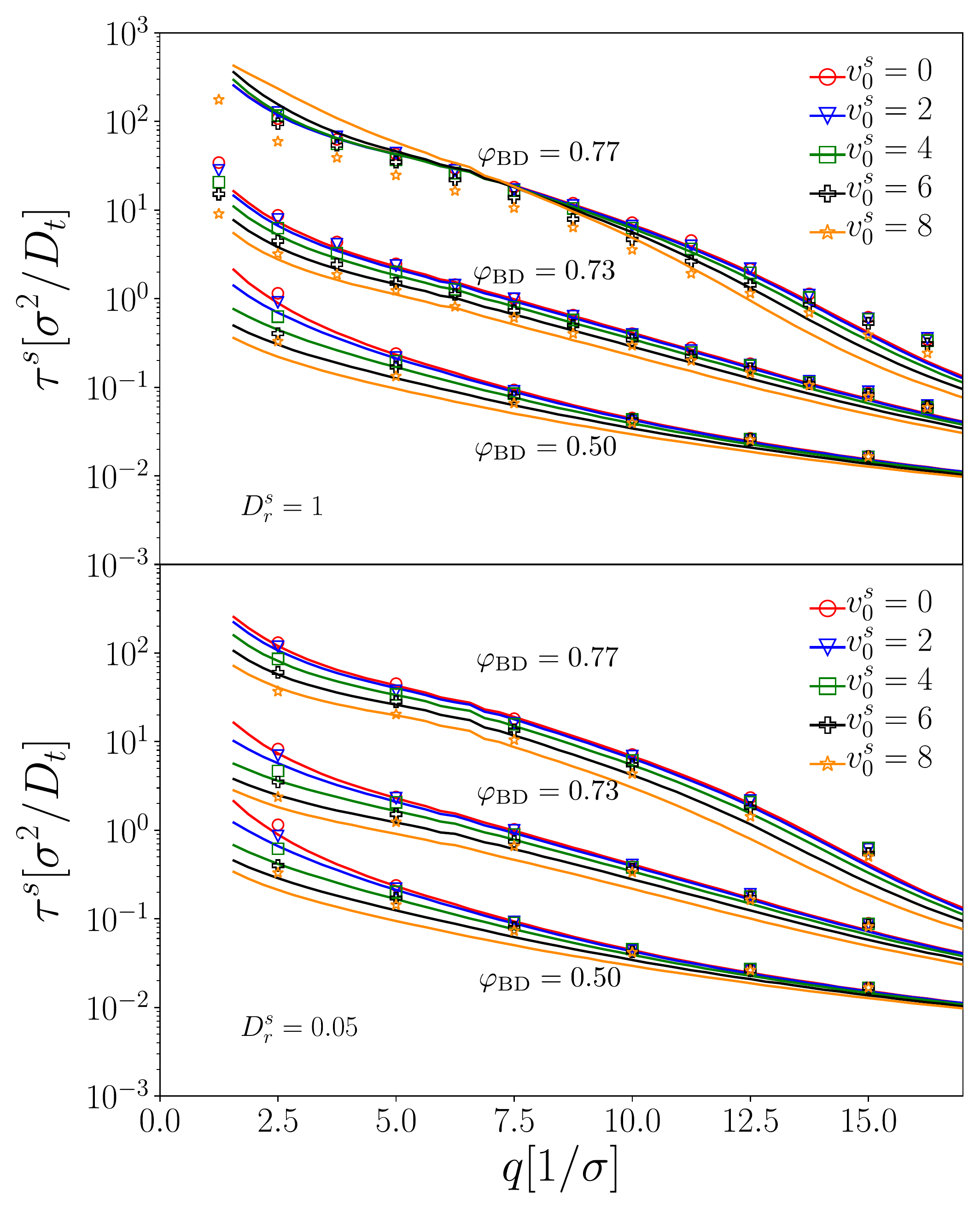}
\caption{\label{fig:active-passive-tau}
  Structural relaxation time $\tau^s$
  of the self-intermediate scattering function of
  an active tracer in a passive host system, for different packing fractions
  $\varphi$ as indicated, and as a function of wave number $q\sigma$.
  Symbols are from ED-BD simulations, lines are from MCT.
  Top panel: $D_r^s=1\,D_t/\sigma^2$, bottom panel $D_r^s=0.05\,D_t/\sigma^2$.
}
\end{figure}

The influence of the tracer-particle activity in a passive host medium
is most pronounced at low $q$. This is highlighted by the $q$-dependent
structural relaxation times $\tau^s(q)$
extracted from the correlation functions,
determined here for simplicity as the time where the functions have decayed to
the value $1/e$.
Overall, \gls{ABPMCT} provides a good description of the simulation data
(Fig.~\ref{fig:active-passive-tau}),
except for the highest density, $\varphi=0.77$, studied in the simulations
in the case of $D_r^s=1$. There, the theory predicts a crossing of the
$\tau^s$-vs-$q$ curves for different $v_0^s$ of which we find no evidence
in the simulations. It has to be kept in mind, that generally, \gls{MCT}
does not accurately account for the relaxation dynamics very close to the
glass transition, and the discrepancy observed for the low-$q$ data at
$\varphi=0.77$ may be due to that.

For the parameters investigated here, no qualitative changes are observed
between the active-tracer relaxation times and the ones obtained for a
passive tracer ($v_0^s=0$ curves in the figure). For $q\to0$, the latter
approaches a $1/q^2$ asymptote whose prefactor is related to the
long-time diffusion coefficient. The appearance of this diffusive scaling
regime is shifted to lower $q$ with increasing tracer activity; the
corresponding prefactor decreases and thus indicates a larger long-time
diffusion coefficient. This is in qualitative agreement with the overall
acceleration of the dynamics due to activity.

Increasing the persistence length of active motion at the same self-propulsion
speed and density is predicted by \gls{ABPMCT} to cause a stronger
enhancement of relaxation; the simulations for $D_r^s=1$ and $D_r^s=0.05$
shown in Fig.~\ref{fig:active-passive-tau} show this effect for the highest
packing fraction and the lower range of $v_0^s$ that we have studied here.

\subsection{Passive Tracer in an Active Suspension}\label{sec:results:passive_in_active}

\begin{figure}
\includegraphics[width=\linewidth]{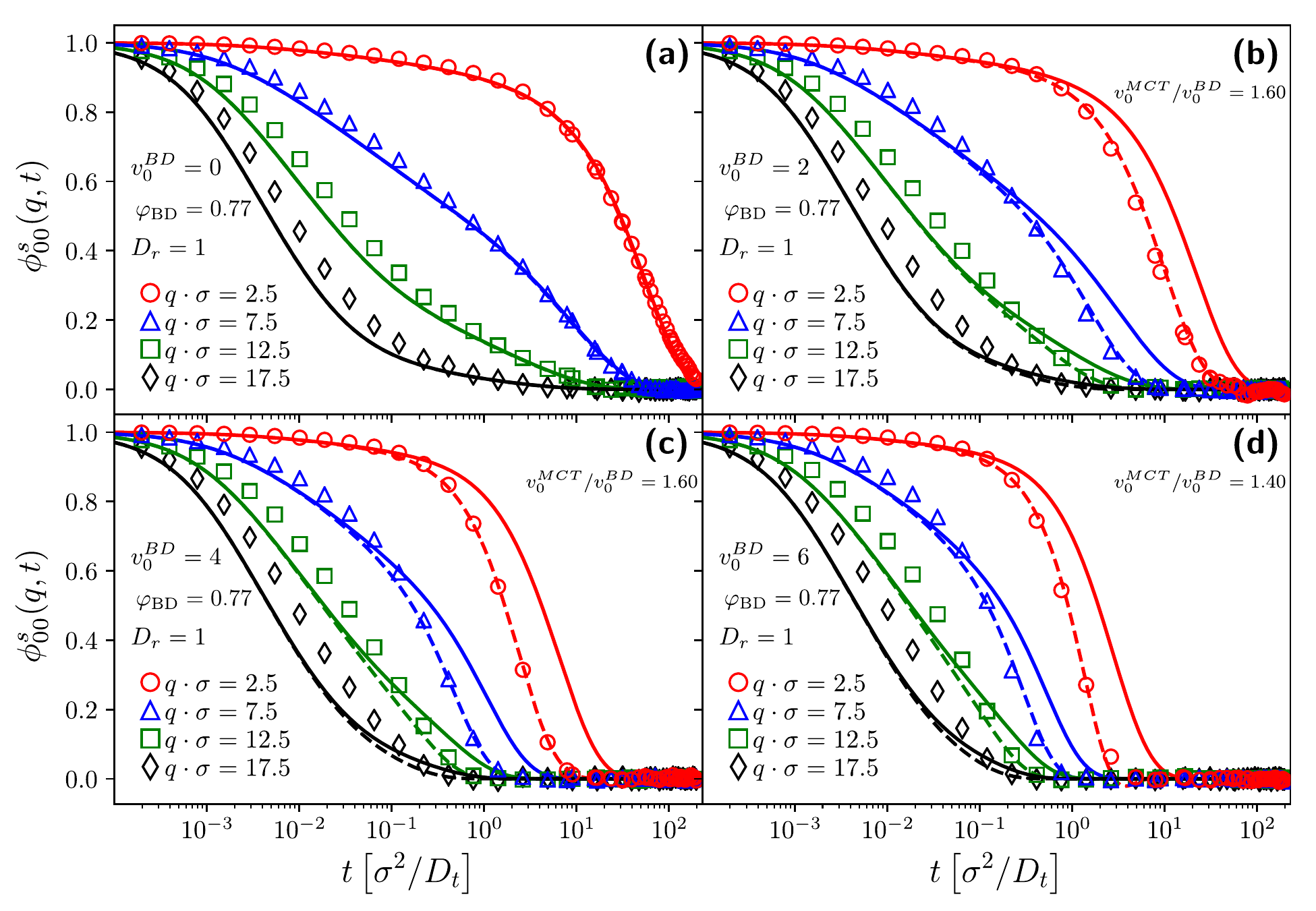}
\caption{\label{fig:passive-active}
  Self-intermediate scattering function of a passive tracer in an active bath
  of packing fraction $\varphi=0.77$.
  Symbols are BD simulation results,
  full lines are from MCT; panels (a)--(d) show various wave numbers as
  indicated for the bath-particle self-propulsion speeds
  $v_0=0$, $2$, $4$, and $6\,D_t/\sigma$.
  Dashed lines represent MCT with a further
  empirical mapping of self-propulsion velocities, $v_0^\text{MCT}\approx
  1.5v_0^\text{BD}$ (as labeled; see text for an explanation).
}
\end{figure}

The passive tracer dynamics in a suspension of ABP provides an interesting
test case for the application of microrheological techniques in active
fluids. Qualitatively, the correlation functions display the same features
as those discussed above for an active tracer in a passive bath
(Fig.~\ref{fig:passive-active}): with increasing activity, the
relaxation time shortens, similarly for all wave numbers.
The effect of an active bath on a passive tracer is, as perhaps might be
expected, much stronger than that of a single active tracer of the same
self-propulsion velocity within a passive bath.

Again, \gls{MCT} is in good qualitative agreement with our BD simulation results.
However, \gls{MCT} underestimates the effect of bath activity on the tracer, i.e.,
the predicted correlation functions decay too slowly.
It is a known effect from other non-equilibrium extensions of \gls{MCT},
that the theory's approximations tend to underestimate the tendency to
fluidization \cite{Henrich.2009,Gazuz.2009}.
This is also in line with the general trend of \gls{MCT} to
overestimate glassiness. Already for sheared colloidal suspensions, an
empirical strain scale could be introduced to remedy this effect \cite{Fuchs.2010}.
Motivated by this finding, we also compare our BD
simulations with the \gls{MCT} predictions where, in addition to the density
mapping that is kept fixed throughout after adjusting it for the purely
passive system, we also introduce an empirical velocity rescaling.
The result (dashed lines in Fig.~\ref{fig:passive-active}) shows that
a roughly parameter-independent mapping of $v_0^\text{MCT}\approx1.5v_0^\text{BD}$
is sufficient to bring \gls{MCT} in semi-quantitative agreement with simulation.

\begin{figure}
\includegraphics[width=\linewidth]{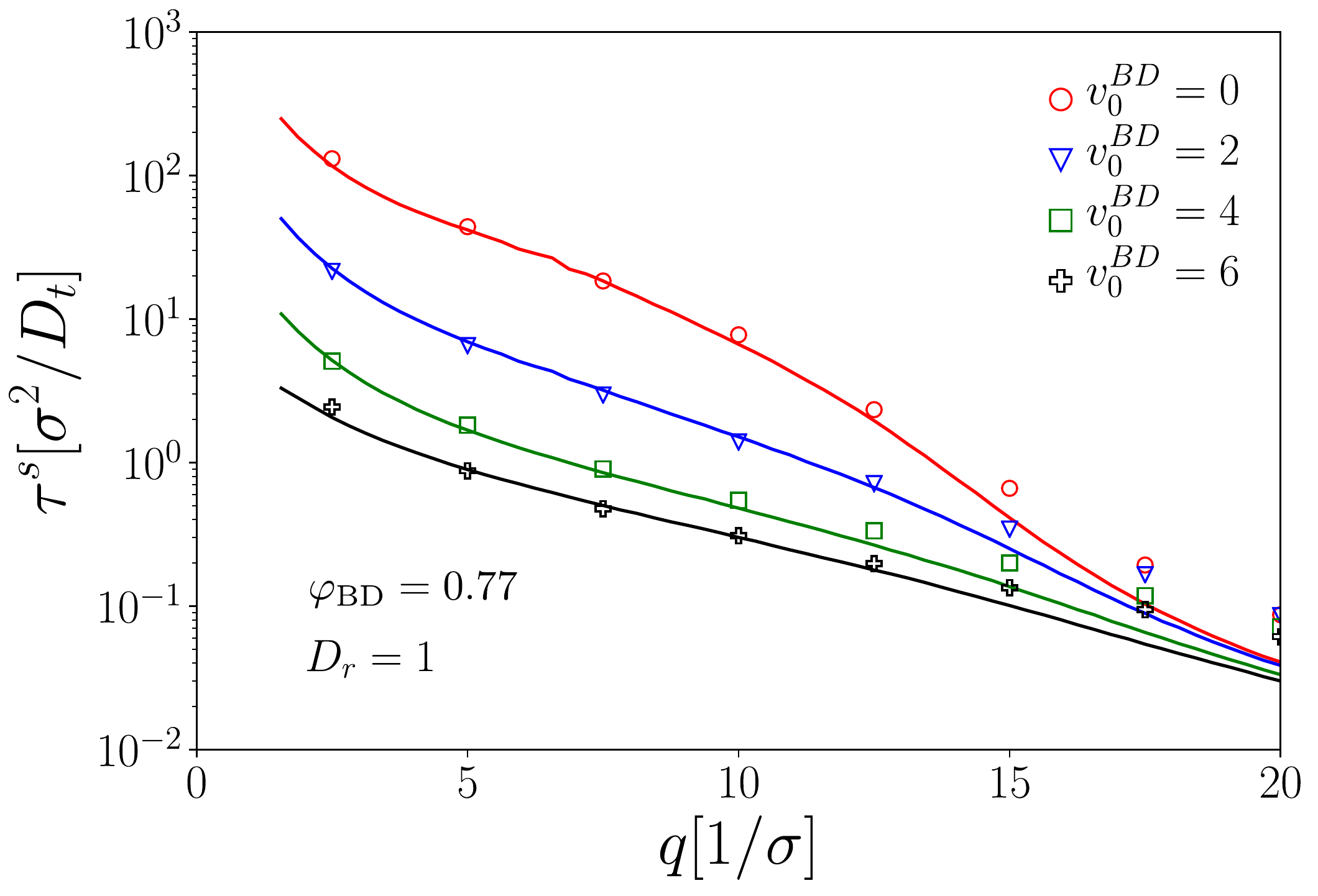}
\caption{\label{fig:passive-active-tau}
  Wave-number dependent
  structural relaxation times $\tau^s(q)$ of a passive tracer in an
  active host system, from ED-BD simulations (symbols) and from
  MCT (lines), at a fixed
  packing fraction $\varphi=0.77$, and various self-propulsion speeds as labeled.
}
\end{figure}

The relaxation time $\tau^s(q)$ for the case of a passive tracer in the
active bath
(Fig.~\ref{fig:passive-active-tau}) demonstrates the much stronger influence
of self-propulsion in this case. Here, the strong influence extends over the
relevant $q$-range in structural relaxation, i.e., up to $q\sigma\approx15$.
One also notes that the hydrodynamic limit, where $\tau^s(q)\sim1/q^2$,
is reached only at $q\sigma\lesssim2$, and there is an intermediate range
of wave numbers where a weaker variation of $\tau^s$ with $q$ becomes
apparent. We shall return to this when discussing the case of an active tracer in an
active host system.

\subsection{Active Tracer in an Active Suspension}\label{sec:results:active_in_active}

\begin{figure}
\includegraphics[width=\linewidth]{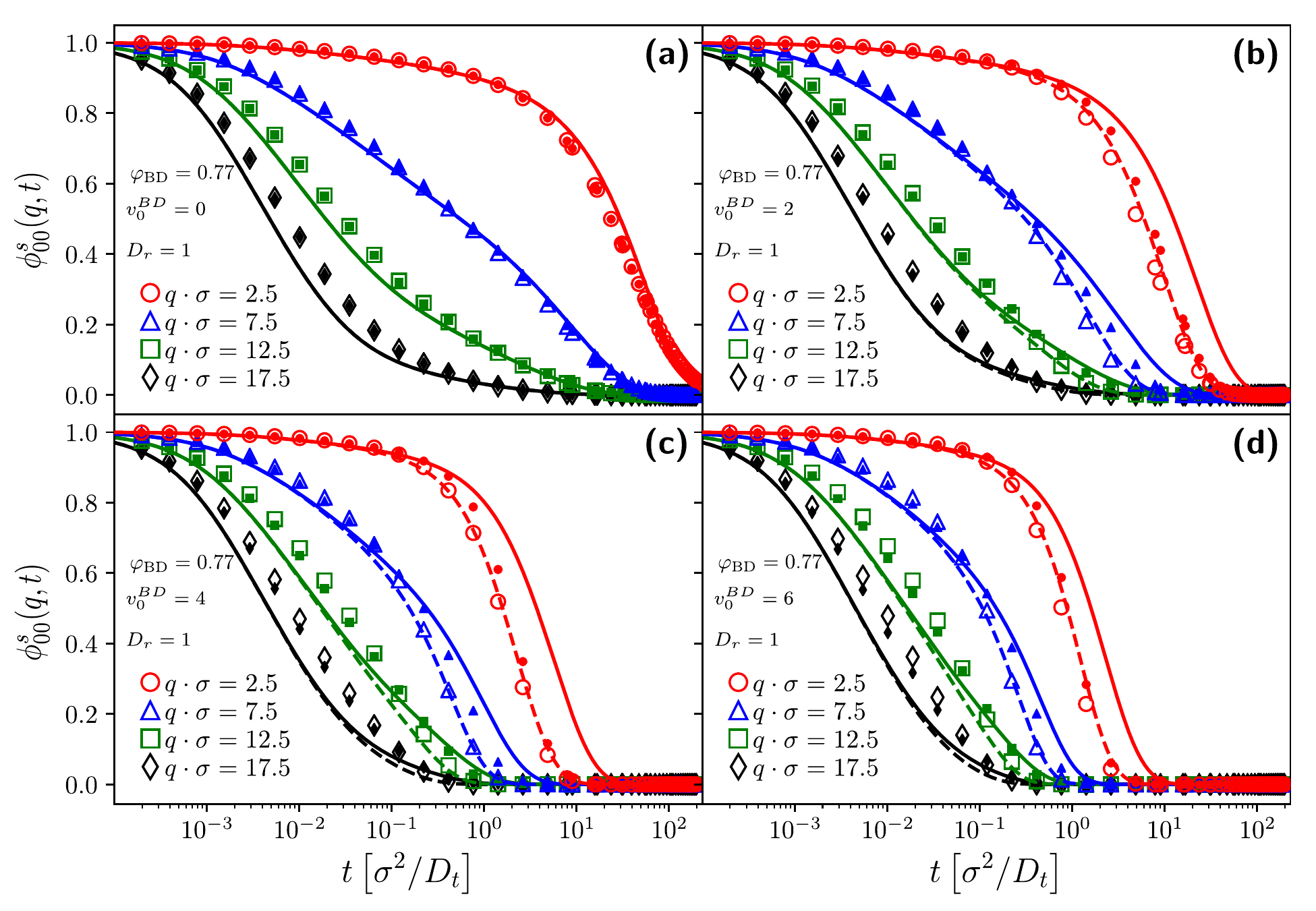}
\caption{\label{fig:active-active}
  Self-intermediate scattering functions of an active tracer in an active bath;
  for fixed bath packing fraction $\varphi=0.77$ and different
  self-propulsion velocities $v_0^s=v_0=0$, $2$, $4$, and $6\,D_t/\sigma$
  [panels (a)--(d)].
  Symbols are results from ED-BD simulations, solid lines are MCT predictions,
  for various wave
  numbers as indicated. Dashed lines show the result of MCT with an adjusted
  self-propulsion velocity $v_0^\text{MCT}=1.5 v_0^\text{BD}$ (see text).
  Open symbols indicate simulation averages over the stationary ensemble,
  while filled symbols represent transient correlation functions defined
  through the passive-equilibrium average involving the full active dynamics.
}
\end{figure}

Finally, the dynamics of an active tracer particle in a fully active
system displays qualitatively similar features as the case of a
passive tracer in the active bath that we just discussed. The density
correlation functions at high densities still decay monotonically,
and their relaxation time decreases with increasing $v_0$
(Fig.~\ref{fig:active-active}).
As in the previous discussion, \gls{MCT} qualitatively accounts for this enhanced
relaxation, but it underestimates its effect. Again, this can be,
for the range of self-propulsion velocities that we consider here,
accounted for by a simple empirical adjustment $v_0^\text{MCT}=1.5 v_0^\text{BD}$,
i.e., by assuming again that \gls{MCT} underestimates the tendency of the
nonequilibrium perturbation to fluidize the system.

\begin{figure}
\includegraphics[width=\linewidth]{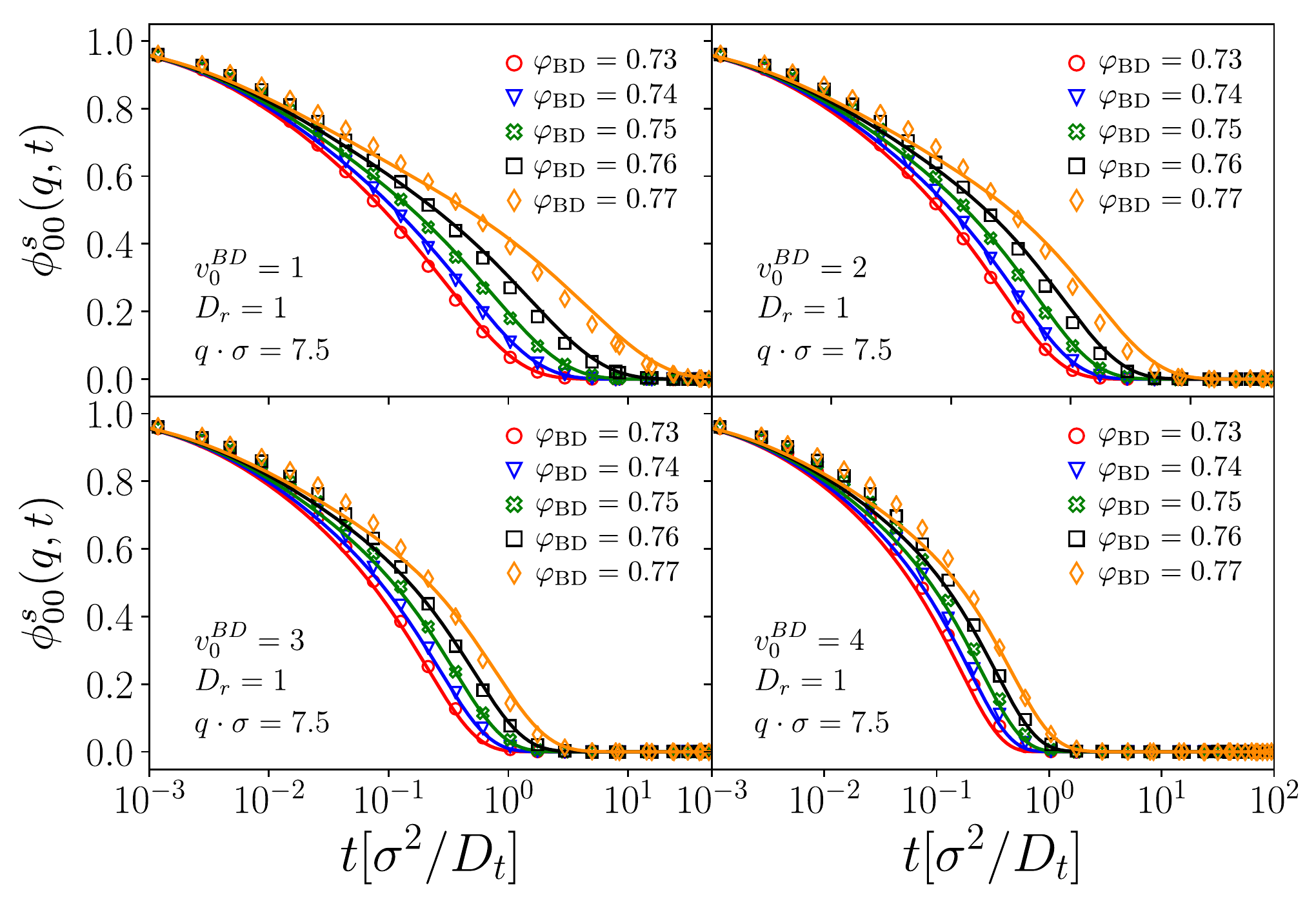}
\caption{\label{fig:active-active-phi}
  Self-intermediate scattering functions for an active tracer in an active
  bath, for self-propulsion velocities $v_0^s=v_0=1$, $2$, $3$, and
  $4\,D_t/\sigma$ [panels (a)--(d)], at fixed wave number $q\sigma=7.5$
  and for different packing fractions $\varphi$ as labeled.
  Symbols are ED-BD simulation results, lines are MCT results with an
  adjusted self-propulsion velocity (see text).
}
\end{figure}

The evolution of the \gls{SISF} with host-system packing fraction in general
resembles that of the approach to a passive glass transition, at least for
the range of self-propulsion speeds that we can compare here with \gls{MCT}
(Fig.~\ref{fig:active-active-phi}). The theory provides an reasonable
quantitative description of the evolution of the dynamics with packing
fraction and self-propulsion velocity close to the glass transition, after
performing only two global adjustments: a
mapping $\varphi_\text{BD}\mapsto\varphi_\text{MCT}$ that is fixed in the
fully passive system and hence not a specific feature of \gls{ABPMCT},
and a mapping $v_0^\text{BD}\mapsto v_0^\text{MCT}$ that accounts for
the stronger effect of activity in simulation compared to theory.

\begin{figure}
\includegraphics[width=\linewidth]{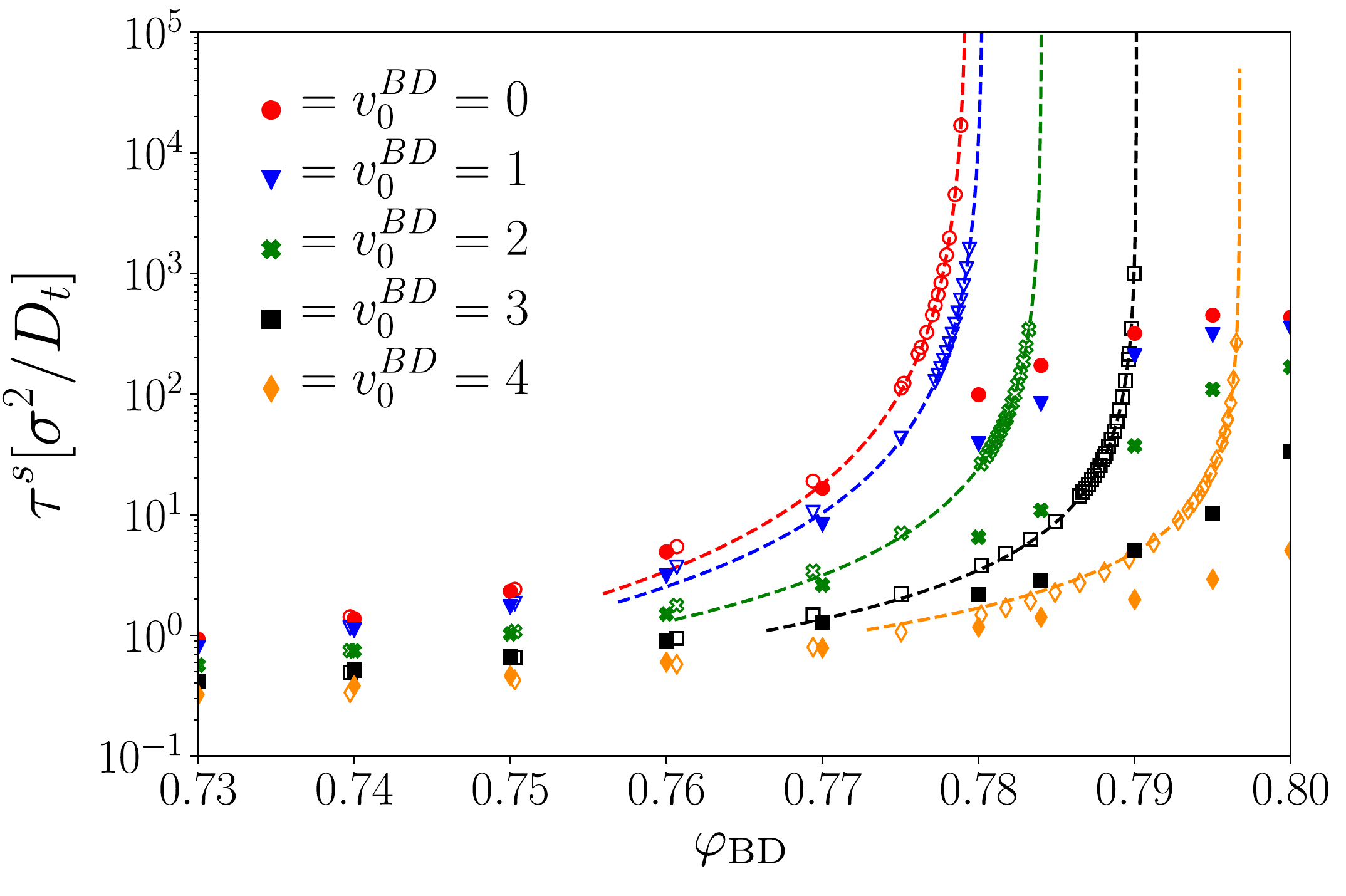}
\caption{\label{fig:active-tauphi}
  Structural relaxation time $\tau^s$ obtained from the \gls{SISF}
  (at wave number $q\sigma=7.5$) of
  an active tracer in an active host system, as a function of packing
  fraction $\varphi$, for various self-propulsion velocities
  $v_0=0\ldots 4\,D_t/\sigma^2$ as labeled. Filled symbols are
  results from ED-BD simulation; numerially obtained relaxation times
  from the solutions of \gls{ABPMCT} are shown as open symbols together
  with corresponding power-law fits (dashed lines).
}
\end{figure}

In the high-density system, the tagged-particle dynamics of a tracer
identical to the bath particles, and the collective dynamics at finite $q$
are strongly coupled, and hence both can be taken as proxies to determine
the structural relaxation time. Indeed, we confirm the expected power-law
divergence of $\tau^s$ within our \gls{ABPMCT}
results for the tagged-particle dynamics
(open symbols and dashed lines in Fig.~\ref{fig:active-tauphi}).
The \gls{EDBD} simulations qualitatiely agree in the window of packing
fractions $\varphi\lesssim0.77$, but then deviate strongly from the predicted
MCT-like divergence. These deviations become more pronounced with increasing
$v_0$. Given that the deviations set in while the relaxation times are
still moderate, $\tau^s\lesssim10^2\,\sigma^2/D_t$, aging effects from
the preparation of the simulation ensemble are unlikely.
In fact, \gls{MCT} is expected to fail close to its predicted glass-transition
point due to the appearance of glass-like relaxation processes that the
theory neglects \cite{Mandal.2018}.
It has been noted previously \cite{Ni.2013} that these deviations appear to
become much stronger in the \gls{ABP} system, so that the
active-glass transition even appears to be shifted up to random-close
packing.
Note however also that we are dealing with a 2D system, where the appearance
of Mermin-Wagner flutuations cause an additional decay channel for the
dynamics as measured by global variables (such as our density correlation
functions) \cite{Illing.2017,Vivek.2017,Shiba.2016}.
A discussion of suitable local variables where the
long-range fluctuations specific to the 2D system cancel out, is beyond the
scope of the present work.

\begin{figure}
\includegraphics[width=\linewidth]{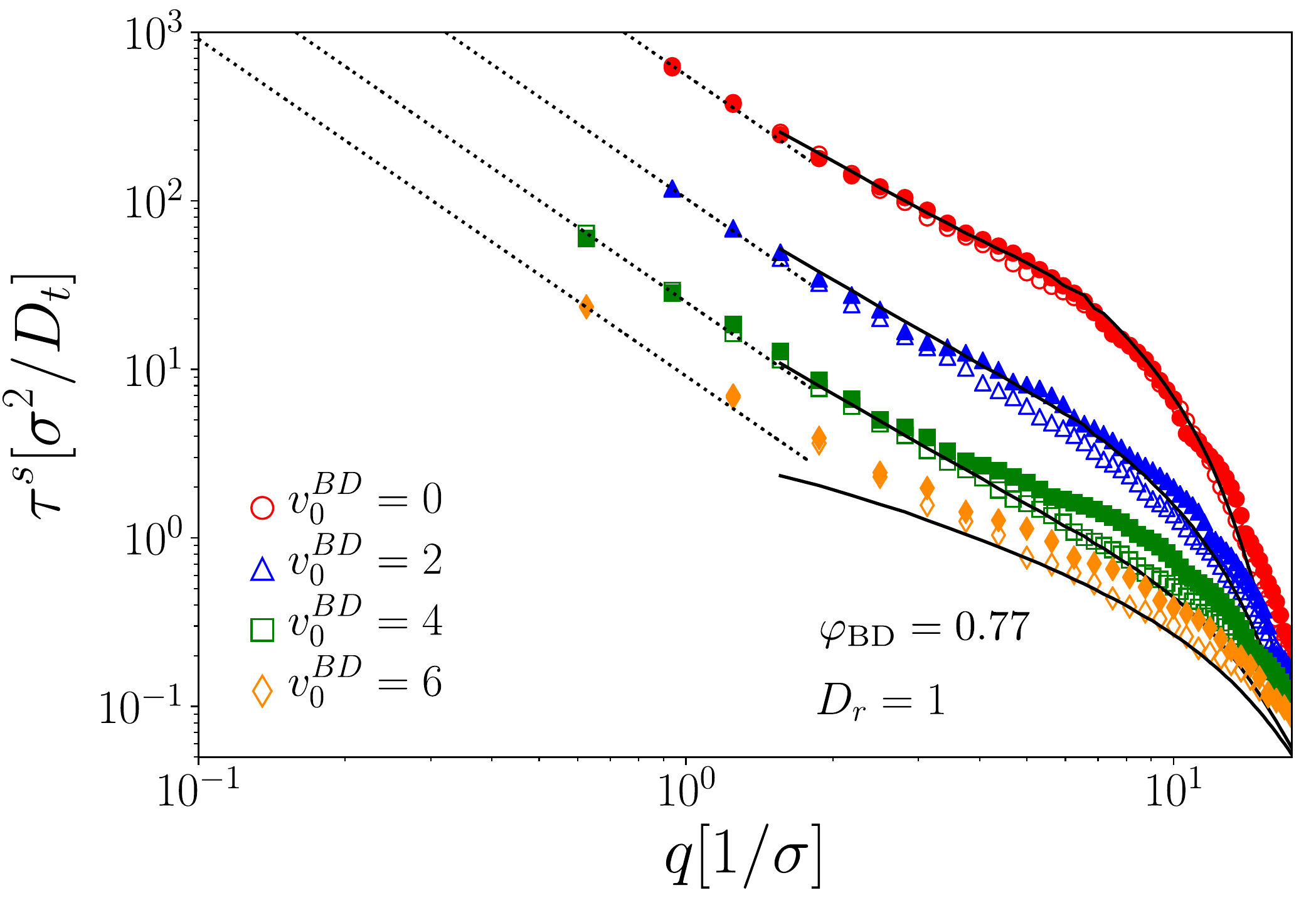}
\caption{\label{fig:active-tauq}
  Wave-number dependence of the structural relaxation time $\tau^s$ of an
  active tracer in an active bath, for packing fraction $\varphi=0.77$
  and different self-propulsion velocities as indicted.
  \gls{ABPMCT} results are shown as solid lines, with $1/q^2$ asymptotes
  for $q\to0$ indicated as dotted lines. \gls{EDBD} simulation results
  are shown as symbols (open: transient; filled: stationary averages).
  For the \gls{ABPMCT} results,
  $v_0^\text{MCT}=1.5v_0^\text{BD}$ has been adjusted.
}
\end{figure}

In the regime where \gls{ABPMCT} still describes the structural relaxation
times, it also predicts the length-scale dependence of structural relaxation
reasonably well.
The $q$-dependence of the relaxation times $\tau^s$ follows, in the regime of high
densities and self-propulsion velocities covered here, the expected behavior
known from passive systems (Fig.~\ref{fig:active-tauq}): as $q\to0$,
a divergence $\tau^s\sim1/q^2$ is observed (dotted lines in the figure).
There then opens a large intermediate regime, where the behavior is
$\tau^s\sim1/q$; note that from the free-particle dynamics and the inversion
of the matrix $\bs\omega^s(\vec q)$ one indeed expects an $1/v_0^sq$ asymptote
to govern the relaxation times [cf.\ Eq.~\eqref{eq:omegaTsinvq0}]. In the \gls{ABPMCT} results for the
highest activity shown in the figure, the wave-number grid that we employ in
the numerical evaluation of the theory precludes the observation of the
$1/q^2$ asymptote in the numerical solutions; it is still seen in the
simulations. This is also expected from the structure of the
matrix $\bs\omega^s(\vec q)$: the onset of the diffusive $1/q^2$ asymptote
is delayed to larger length scales upon increasing the persistence length
of the active motion by increasing $v_0$ at fixed other parameters.

\subsection{Transient \textit{vs.} Stationary Averages}\label{sec:results:transient}

In the previous discussion, we have compared simulation results for the
\gls{SISF} that were obtained under stationary averages, i.e.,
effectively $\phi^{s,\text{stat}}_{ll'}(\vec q,t)=\langle\varrho_l^{s*}
\exp[\smol^\dagger t]\varrho_{l'}^s(\vec q)\rangle_\text{stat}$,
with the transient quantity $\phi^s_{ll'}(\vec q,t)$ defined
in Eq.~\eqref{eq:phis}, for which \gls{ABPMCT} is formulated.
This introduced the tacit assumption that the qualitative (and partially
quantitative) features of the structural-relaxation dynamics are identical
in both quantities.
The difference between the two correlation functions is indeed difficult to
estimate, since the stationary distribution function of
the system is not known.

In the case of an active tracer partice moving in an active host system,
discussed in the previous subsection, our simulations allow to address
this question numerically. In Fig.~\ref{fig:active-active},
both the transient and the stationary \gls{SISF} are shown (filled and
open symbols) for a typical case close to dynamical arrest, i.e., at
large packing fraction and various intermediate wave numbers.
Both quantities follow each other closely over the parameter regime
investigated in that figure.
This confirms the approach to compare solutions of \gls{ABPMCT} with
stationary solutions from simulation or experiment under the premise that
the errors inherent in the theory are typically larger than the differences
in the different correlation functions.
In the \gls{ITT}-\gls{MCT} approach to sheared colloidal suspensions,
a similar statement holds \cite{Krueger.2010}, although there, the transient
correlation functions show a regime of faster-than-exponential decay
that is absent in the stationary ones. No such qualitative differences
are seen in the regime investigated in Fig.~\ref{fig:active-active}.
Also the structural relaxation times obtained from the two different
correlation functions (open and filled symbols in Fig.~\ref{fig:active-tauq})
are qualitatively similar.

\begin{figure}
\includegraphics[width=\linewidth]{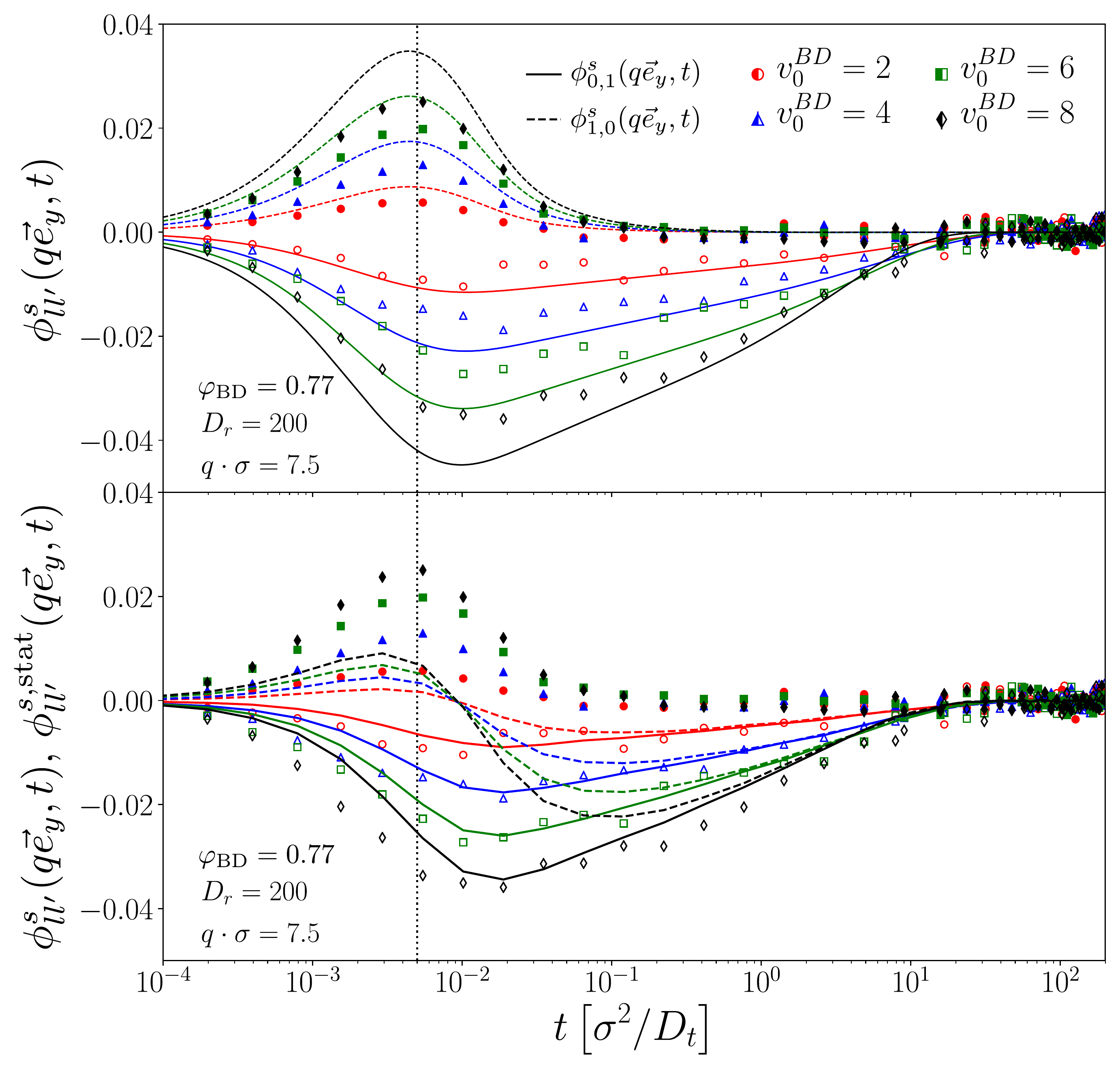}
\caption{\label{fig:active-asym}
  Stationary and transient tagged-particle
  correlation functions $\phi^s_{l,l'}(\vec q,t)$ of the dense ABP system
  (active tracer in active host system; packing fraction $\varphi=0.77$) with $D_r=200\,D_t/\sigma^2$,
  for $\vec q\sigma=7.5\vec e_y$ and self-propulsion velocities
  $v_0=2$, $4$, $6$, and $8\,D_t/\sigma$ as indicated.
  Top panel: transient correlation functions from ED-BD simulations.
  Open symbols show $(l,l')=(0,1)$, filled symbols $(l,l')=(1,0)$.
  Solid and dashed lines are the corresponding ABP-MCT results.
  Bottom panel: comparison of transient (symbols) and stationary (lines)
  correlation functions both from ED-BD simulations. Here, open/filled
  symbols are transient correlators for $(l,l')=(0,1)$ and $(l,l')=(1,0)$
  (as above), and solid/dashed lines the corresponding stationary ones.
  A vertical dotted line indicates the time scale $1/D_r$.
}
\end{figure}

In Brownian dynamics that obeys detailed balance, the backward
Smoluchowski operator $\smol^\dagger$ driving the time evolution,
is self-adjoint in the scalar product weighted with the corresponding
equilibrium distribution function: there holds
$\langle f^*\smol_\text{eq}^\dagger g\rangle_\text{eq}
=\langle (\smol_\text{eq}^\dagger f^*)g\rangle_\text{eq}$
for arbitrary observables $f$ and $g$.
The active-particle dynamics does not have the property of detailed balance,
and hence $\smol^\dagger$ is no longer self-adjoint.
For the correlation functions with angular-mode indices $l,l'$, this
amounts to a breaking of symmetry: under the equilibrium dynamics,
there holds $\bs\Phi(\vec q,t)\stackrel{\text{eq}}=\bs\Phi^T(\vec q,t)$, while this
property is \textit{a~priori} lost in the non-equilibrium dynamics.
In fact, this is already borne out by the analytical low-density solution
(cf.\ Fig.~\ref{fig:lowdens}).

In \gls{ABPMCT}, the resulting asymmetry extends in a peculiar way to the long-time limits
of the correlations functions in the ideal glass, the (tagged-particle)
non-ergodicity
parameters $f_{ll'}^{(s)}$: the feature of the \gls{ABP}
model that rotations always remain ergodic, causes the $(l0)$ matrix
elements of the correlation functions to decay on a time scale $1/D_r$ for
$l\neq0$, while only the $(0l')$ matrix elements pick up slow
structural relaxation on time scales $t\gg1/D_r$ and potentially show
nonergodic arrest. Technically, this arises due to the presence of
the $\bs\omega_R^{(s)}$ contribution in the memory integral in
Eqs.~\eqref{eq:mzphi} and \eqref{eq:mcts}; taking the Laplace transform
of these equations, one verifies that the zero-eigenmodes of
$\bs\omega_R^{(s)}$ always decay exponentially at long times
\cite{Liluashvili.2017}. Intuitively, this expresses that the current
configuration of orientational $l\neq0$ fluctuations are not correlated to how the
positional $l'=0$ fluctuations have evolved (for example, $f^{(s)}_{10}=0$),
while inversely, the
current positional $l=0$ fluctuations are influenced by the distant past
evolution of the $l'\neq0$ orientations through the active driving term
($f^{(s)}_{01}\neq0$).
This still is a subtle prediction, because it hinges on the peculiar structure
of the evolution equations of \gls{ABPMCT}, and on the definition of the
correlation function. (Recall that in the theory of stochastic processes,
it is customary to consider correlation functions that in any stationary
process are symmetric \cite{FellerII}.)

We check this asymmetry in our \gls{EDBD} simulations (Fig.~\ref{fig:active-asym}). To emphasize the difference, we choose $D_r=200\,D_t/\sigma^2$ such
that the time scale $1/D_r$ is much shorter than the structural relaxation
time of the system. Indeed, we observe a notable asymmetry in the decay
pattern of $\phi^s_{1,0}(\vec q,t)$ and $\phi^s_{0,1}(\vec q,t)$ in the
transient correlation functions (top panel of Fig.~\ref{fig:active-asym}).
The \gls{ABPMCT} results are in very good qualitative agreement regarding
this asymmetry. Note that here we have chosen $\vec q\parallel\vec e_y$
such that both correlation functions are real-valued and hence would be
identical in equilibrium dynamics. There are thus two remarkable features
visible in the plot: first, as predicted by \gls{ABPMCT}, the correlation
function for $(l,l')=(1,0)$ decays on the time scale $1/D_r$, where it
induces an incomplete relaxation in the $(l,l')=(0,1)$ correlator. The latter
in turn remains non-zero and shows indications of structural relaxation,
only decaying on the time scale $\tau^s$ (around $1$ to $10\,\sigma^2/D_t$
for the state points considered here). It is plausible to presume that
for higher packing fraction, the quantity $\phi^s_{0,1}(\vec q,t)$ will
(in the simulationally unattainable ideal case)
show structural arrest and thus
a finite long-time limit, while $\phi^s_{1,0}(\vec q,t)$ clearly
decays to zero. Second, the two correlation functions display opposite
signs; a feature already encoded in the short-time dynamics of the \gls{ABP}.

Within the simulation, we also check whether the asymmetry still holds
for the stationary correlation functions (lines in the lower panel of
Fig.~\ref{fig:active-asym}). Interestingly, a strong asymmetry remains
on time scales small compared to $1/D_r$, while for large times,
$t\gg1/D_r$, the two \emph{stationary} correlation functions approach
each other and show structural relaxation that is identical.
Thus, extending \textit{bona fide} the \gls{ABPMCT} scenario, we conjecture
that in the ideal-arrest case, the \emph{stationary} non-ergodicity parameters
will all be non-vanishing and form a symmetric matrix,
$f^{(s),\text{stat}}_{ll'}=f^{(s),\text{stat}}_{l'l}\neq0$,
by a non-trivial coupling of orientational and translational fluctuations
through the stationary distribution function.

\section{Conclusion}\label{sec:conclusion}

We have derived equations of motion for the angle-resolved tagged-particle density
correlation functions of tracers in dense systems of active Brownian particles,
in the context of the mode-coupling theory of the glass transition and
the integration-through transients framework.
The numerical solutions of this \gls{ABPMCT} theory have been compared
to computer simulations of hard-disk \gls{ABP} for the different
cases of active and passive tracers in active and passive host systems.
Overall, we have found theory and simulation to be in good qualitative
agreement, for a range of densities close to the glass transition,
and for a range of self-propulsion speeds that is not too large.
The theory, as known for the passive \gls{MCT}, overestimates the
slowness of structural relaxation, and hence the comparison to
simulation data best proceeds with an empirical mapping of densities,
and an empirical
mapping of self-propulsion speeds in the case of an active host system.
After such mapping, the agreement between theory and simulation is
near-quantitative.

An interesting feature in the angle-resolved correlation functions is that
they are in general non-symmetric matrices of the angular-mode indices.
This reflects the non-equilibrium nature of the dynamics, and has first
been noted in the \gls{ABPMCT} for the collective density fluctuations.
We have confirmed this asymmetry directly in our \gls{BD} simulations.
The simulations point out an interesting difference in the transient
correlation functions targeted by the \gls{ITT}-\gls{MCT}, and the
stationary correlation functions more commonly accessed in experiment
and simulation: while both remain asymmetric, there appears to be a
time scale after which the time-evolution crosses over to a symmetric
form of correlations (as perhaps more intuitively expected).
The time scale is associated with the rotational persistence of the
\gls{ABP}; it is thus expected to also play an important role in
determining, for example, signatures of persistant motion in other
quantities such as the \gls{MSD} (where super-diffusive regimes indicate
the time windows of persistant swiming). Such investigation will be
discussed elsewhere.

\begin{acknowledgments}
We acknowledge funding from Deutsche Forschungsgemeinschaft (DFG), as part of the Special Priority Programme SPP~1726 ``Microswimmers'' (project Vo~1270/7).
\end{acknowledgments}

\begin{appendix}

\section{Inversion of Frequency Matrix}\label{sec:omegainv}

A tridiagonal $n\times n$
matrix with elements $a_{ij}$ such that $a_{ij}=0$ whenever
$|i-j|>1$, can be inverted by employing a recursion formula for the
elements $\alpha_{ij}$ of the inverse matrix. Setting
$a_{ii}=b_i$, $a_{i,i+1}=c_i$, and $a_{i,i-1}=c_i$, Usmani \cite{Usmani.1994}
derives
\begin{equation}
  \alpha_{ij}=\begin{cases}
    (-)^{i+j}c_ic_{i+1}\cdots c_{j-1}\theta_{i-1}\phi_{j+1}/\theta_n\,,
    & i<j,\\
  \theta_{i-1}\phi_{i+1}/\theta_n\,, & i=j,\\
  (-)^{i+j}a_{j+1}a_{j+2}\cdots a_i\theta_{j-1}\phi_{i+1}/\theta_n\,,
    &i>j,
  \end{cases}
\end{equation}
where the sequences $\theta_i$ and $\phi_i$ are given by recursion relations,
\begin{subequations}
\begin{align}
  \theta_i&=b_i\theta_{i-1}-a_ic_{i-1}\theta_{i-2}\,,\\
  \phi_i&=b_i\phi_{i+1}-c_ia_{i+1}\phi_{i+2}\,,
\end{align}
\end{subequations}
with $\theta_0=\phi_{n+1}=1$ and $\theta_{-1}=\phi_{n+2}=0$.

We apply this result for the $n\times n$ matrix $\bs\omega_T^s$ given
by Eq.~\eqref{eq:omegaTs} with
cutoff $L$, setting $n=2L+1$, $i=l+L+1$ and $j=l'+L+1$ with
$l,l'\in[-L,L]$. Identifying then $b_i=D^s_tq^2$, $a_i=c_i=-iv_0^sq/2$,
the recursion relations can be solved in closed form: setting
$\Delta=\sqrt{(D^s_tq^2)^2+(v_0^sq)^2}$, and $\beta_\pm=D^s_tq^2\pm\Delta$,
one obtains
\begin{subequations}
\begin{gather}
  \theta_i/\theta_n=\frac{2^{n-i}\left(\beta_+^{1+i}-\beta_-^{1+i}\right)}
  {\beta_+^{1+n}-\beta_-^{1+n}}\,, \\
  \begin{split}
  \phi_i=\frac{(-)^{i+1-n}}\Delta
  2^{i-2-n}(v_0^sq)^{2(n-i)}\times \\
  \times \Bigl(\beta_+^{i-n}((v_0^sq)^2+2D^s_tq^2\beta_-)
  \\
  -\beta_-^{i-n}((v_0^sq)^2+2D^s_tq^2\beta_+)\Bigr)
  \end{split}
\end{gather}
\end{subequations}
From this, the limit $L\to\infty$ at fixed $l,l'$ and fixed $q$ can be performed. Using $\lim_{L\to\infty}(\beta_-/\beta_+)^L=0$, one finally gets
Eq.~\eqref{eq:omegaTsinv}.

\section{Low-$q$ Expansion of the Memory Kernel}\label{sec:lowqdetail}

The limit $q\to0$ in the vertex, Eq.~\eqref{eq:Ws} is obtained noting
$k=p-(\vec q\cdot\vec p)/p+\mathcal O(q^2/p)$ and
$k\exp[\pm i\theta_k]=q\exp[\pm i\theta_q]-p\exp[\pm i\theta_p]$ for
vectors $\vec q=\vec k+\vec p$.
Note that the equilibrium contribution is already of $\mathcal O(q^2)$,
so that in this case $\tilde{\bs m}^s(q,t)=\mathcal O(q^0)$, and
hence the $(00)$ element of Eq.~\eqref{eq:mcts} is $\mathcal O(q^2)$
in leading order, compatible with the expression
Eq.~\eqref{eq:phismsd}.

This is less obvious for the case $v_0^s\neq0$ or $v_0\neq0$. In particular,
the low-$q$ expansion of Eq.~\eqref{eq:Ws} contains terms of $\mathcal O(q)$.
One needs to check that these potentially dangerous terms do not
contribute in the Mori-Zwanzig equation. Inserting the small-$q$ expansion
of $\mathcal W^s(\vec q,\vec k)$ into Eq.~\eqref{eq:Msmct} and performing
the integration over the angle $\theta_k$ in $\int d\vec k=\int k\,dk\,d\theta_k$, one obtains
\begin{multline}\label{eq:Ms0}
  \tilde M_{ll'}^s(q,t)\sim\frac{n{D_t^s}^2}{8\pi}\int dp\,
  \tilde\phi^s_{m'l'}(p,t)\times\\
  \Bigl\{q^2p^3(\delta_{l-l',-2}+\delta_{l-l',2}+2\delta_{ll'})
  c^s(p)^2\Phi_{00}(p,t)\delta_{lm'}\\
  +\frac{iqp^3}{D_t^s}(\delta_{l-l',-1}+\delta_{l-l',1})
  \tilde w_{l,mm'}(p)\tilde\Phi_{m0}(p,t)\\
  -\frac{iq^2}{D_t^s}\delta_{ll'}\partial_p\left(p^3\tilde w_{l,mm'}(p)
  \tilde\Phi_{m0}(p,t)\right)
  \\
  -\frac{iq^2p^2}{D_t^s}(\delta_{l-l',-2}+\delta_{l-l',2})
  \partial_p\left(p\tilde w_{l,mm'}(p)\tilde\Phi_{m0}(p,t)\right)
  \Bigr\}
\end{multline}
where we have defined
\begin{equation}
  \tilde w_{l,mm'}(p)=\left(\delta_{|l-m'|,1}\delta_{m0}v_0^s
  -\delta_{lm'}\delta_{|m|,1}v_0S(p)\right)c^s(p)^2\,.
\end{equation}

For the passive tracer, $v_0^s=0$, the term $\tilde w_{l,mm'}(p)$ is
proportional to $\delta_{lm'}$, so that all terms appearing in the
memory kernel, Eq.~\eqref{eq:Ms0},
couple to $\tilde\phi^s_{ll'}(p,t)$. But since for this
case, the matrix $\tilde{\bs\omega}_T^s(q)$ remains diagonal,
the Mori-Zwanzig equations that determine the off-diagonal elements of the
correlation-function matrix decouple and, since
$\tilde\phi^s_{ll'}(q,0)=\delta_{ll'}$, these off-diagonal elements all
identically vanish. The only contribution to $\tilde m^s_{ll'}(q,t)$
is thus from the diagonal elements, all of $\mathcal O(q^2)$ in leading
order. One obtains the only memory kernel that enters
the \gls{MSD} equation, Eq.~\eqref{eq:msd}, for the passive tracer:
\begin{multline}\label{eq:mtildes0passive}
  \hat m^s_{00}(t)=\frac{nD_t^s}{4\pi}\int dp\,\tilde\phi^s_{00}(p,t)
  \Bigl\{p^3c^s(p)^2\tilde\Phi_{00}(p,t)\\
  +\frac{i}{2D_t^s}\delta_{|m|,1}\partial_pp^3v_0S(p)c^s(p)^2\tilde\Phi_{m0}(p,t)
  \Bigr\}\,.
\end{multline}
The first term is the known contribution from passive \gls{MCT}
\cite{Bayer.2007}.

In the case of an active tracer particle, Eq.~\eqref{eq:Ms0} is
simplified by noting a further symmetry of the correlation functions:
inverting the coordinate system around the $x$-axis, $\{x_k,y_k,\varphi_k\}\mapsto\{x_k,-y_k,-\varphi_k\}$, invariance of both the equilibrium distribution
and the Smoluchowski operator implies that
$\tilde\Phi_{l,l'}(q,t)=\tilde\Phi_{-l,-l'}(q,t)$. One further checks
that $\tilde\phi^s_{l0}(q,t)=\mathcal O((iqv_0^s)^{|l|})$ from checking
explicitly the terms $(\smol^\dagger)^n\exp[i\vec q\cdot\vec r_s]$ stemming
from the expansion of $\exp[\smol^\dagger t]$, and
observing that the non-vanishing angular integrals in $\tilde\phi^s_{l0}(q,t)$
require at least the $l$-fold application of $\dsmol^\dagger$.

The low-$q$ expansion of the memory kernel, Eq.~\eqref{eq:Ms0}, then needs
to be combined with the low-$q$ expansion of $\bs\omega_T^{s,-1}(q)$,
Eq.~\eqref{eq:omegaTsinvq0}, in order to check the leading contributions
for the different angular indices.
The memory kernels that are
relevant for the \gls{MSD} then read
\begin{multline}\label{eq:mtildes0active00}
  \hat m^s_{00}(t)=\frac{nD_t^s}{8\pi}\frac1{v_0^s}\int dp\,\tilde\phi^s_{m'l'}(p,t)\times\\ \left\{
  \delta_{|l'|,1}p^3\tilde w_{0,mm'}(p)\tilde\Phi_{m0}(p,t)\right\}
\end{multline}
and, determining the coupling to the off-diagonal correlator
$\hat\phi^s_{\pm1,0}(t)$,
\begin{multline}\label{eq:mtildes0active01}
  \hat m^s_{0,\pm1}(t)=\frac{nD_t^s}{4\pi}\frac1{v_0^s}
  \int dp\,\tilde\phi^s_{m'l'}(p,t)\times\\
  \Bigl\{iD_t^sp^3c^s(p)^2\tilde\Phi_{00}(p,t)\delta_{m'0}\delta_{l'0}
  \\
  -\frac12\partial_pp^3\tilde w_{0,mm'}(p)\tilde\Phi_{m0}(p,t)\delta_{l'0}\\
  -\frac{D_t^s}{v_0^s}i\delta_{|l'|,1}p^3\tilde w_{0,mm'}(p)\tilde\Phi_{m0}(p,t)
  \Bigr\}\,.
\end{multline}
with a memory kernel
\begin{multline}\label{eq:mtildes0active1}
  \hat m^s_{\pm1,\pm1}(t)=\frac{nD_t^s}{8\pi}\frac1{v_0^s}
  \int dp\,\delta_{|1-l'|,1}\tilde\phi^s_{m'l'}(p,t)\times\\ p^3\tilde w_{1,mm'}(p)\tilde\Phi_{m0}(p,t)\,.
\end{multline}

\section{Mathieu Function Representation}\label{sec:mathieu}

As discussed by Kurzthaler et~al.\ \cite{Kurzthaler.2018,Kurzthaler.2016}, the
density-correlation function $\phi_{ll'}(\vec q,t)$
of a free \gls{ABP} can be related to the
differential equation familiar from the solution of the
Schr\"odinger equation with a periodic potential: One writes,
generalizing Ref.~\cite{Kurzthaler.2018} to $(ll')\neq(00)$,
\begin{multline}\label{eq:mathieuphi}
  \phi_{ll'}(\vec q,t)\\ =\int d\varphi\,d\varphi_0\,d\vec r\,d\vec r_0\,
  e^{il'\varphi}e^{-il\varphi_0}e^{i\vec q\cdot(\vec r-\vec r_0)}
  p(\vec r,\varphi,t|\vec r_0,\varphi_0)\\
  =\int d\varphi\,d\varphi_0\,e^{il'\varphi}e^{-il\varphi_0}
  \tilde p(\vec q,\varphi,t|\varphi_0)\,,
\end{multline}
where we have used that due to translational symmetry the initial position
$\vec r_0$ can be integrated out. The characteristic function $\tilde p$
obeys
\begin{equation}\label{eq:mathieuptilde}
  \partial_t\tilde p=\left[-D_tq^2+iv_0\vec n(\varphi)\cdot\vec q
  +D_r\partial_\varphi^2\right]\tilde p\,,
\end{equation}
where it will be convenient to choose $\vec q\parallel\vec e_x$.
This differential equation is separable, and the angular part reduces
(in two spatial dimensions) to
the Mathieu equation
\begin{equation}\label{eq:mathieudgl}
  \left[\frac{d^2}{dx^2}-2k\cos(2x)\right]f(x)=\lambda(k)f(x)\,,
\end{equation}
after substitution $x=\varphi/2$ and $2k=-iv_0q/D_r$. The solutions of
Eq.~\eqref{eq:mathieudgl} are the even and odd Mathieu functions
$\ce_{2n}(k,x)$ and $\se_{2n+1}(k,x)$ with eigenvalues
$\lambda(k)=a_{2n}(k)$ and $b_{2n+1}(k)$, respectively
\cite{Ziener.2012}. They each form an orthonormal basis and can be expressed
through their cosine and sine series,
\begin{subequations}\label{eq:mathieuexpand}
\begin{align}
  \ce_{2n}(k,x)&=\sum_{m=0}^\infty A_{2m}^{2n}(k)\cos(2mx)\,,\\
  \se_{2n+1}(k,x)&=\sum_{m=0}^\infty B_{2m}^{2n+1}(k)\sin(2(m+1)x)\,,
\end{align}
\end{subequations}
with orthogonality relations for the expansion coefficients
\begin{subequations}\label{eq:mathieuAB}
\begin{align}
  \sum\nolimits_nA_{2m}^{2n}A_{2m'}^{2n}&=\delta_{mm'}-\frac12\delta_{m0}\delta_{m'0}\,,
  \\
  \sum\nolimits_nB_{2m}^{2n+1}B_{2m'}^{2n+1}&=\delta_{mm'}\,.
\end{align}
\end{subequations}
Inserting Eqs.~\eqref{eq:mathieuexpand} in
Eq.~\eqref{eq:mathieuptilde}, one can express $\phi_{ll'}(\vec q,t)$
in Eq.~\eqref{eq:mathieuphi} through its series expansion. For this, we make
use of
\begin{align}
  \int_0^{2\pi}\frac{d\varphi}{2\pi}\ce_{2n}(k,\varphi/2)e^{\pm il\varphi}&=\frac12\left(A_{|2l|}^{2n}+A_0^{2n}\delta_{l0}\right)\,,\\
  \int_0^{2\pi}\frac{d\varphi}{2\pi}\se_{2n+1}(k,\varphi/2)e^{\pm il\varphi}&=\pm\frac{i\sgn l}{2}B_{|2l|-2}^{2n+1}\,.
\end{align}
We obtain
\begin{multline}\label{eq:mathieuseries}
  \phi_{ll'}(q\vec e_x,t)=e^{-q^2D_tt}\sum_{n=0}^\infty\Bigl\{\\
  \frac12\left(A_{|2l'|}^{2n}+A_0^{2n}\delta_{l'0}\right)
         \left(A_{|2l|}^{2n}+A_0^{2n}\delta_{l0}\right)
  e^{-D_ra_{2n}(k)t/4}\\
  -\frac i2(\sgn l)(\sgn l')B_{|2l'|-2}^{2n+1}B_{|2l|-2}^{2n+1}
  e^{-D_rb_{2n}(k)t/4}\Bigr\}\,.
\end{multline}
In the special case of the \gls{SISF}, $l=l'=0$, only the even expansion
terms proportional to $[A_0^{2n}(k)]^2$ remain. Our expression then
coincides with Eq.~(20) of the Supplementary Material of
Ref.~\cite{Kurzthaler.2018}.

Equation~\eqref{eq:mathieuseries} can be readily rewritten as $\phi_{ll'}(q\vec e_x,t)
=\sum_n\mathcal T_{ln}\exp[-\mathcal D_{nn}t]\mathcal U_{nl'}$, identifying
$\bs{\mathcal U}=\bs{\mathcal T}^{-1}$ with the help of the orthogonality
relations Eq.~\eqref{eq:mathieuAB} (see Ref.~\cite{ReichertPhD}),
and identifying the diagonal matrix
\begin{equation}
  \mathcal D_{ll}=\begin{cases}D_tq^2+D_ra_{2n}(k)/4 & \text{for $|l|=2n$,}\\
  D_tq^2+D_rb_{2n+1}(k)/4 & \text{for $|l|=2n+1$.}\end{cases}
\end{equation}
Thus, $\bs\phi(q\vec e_x,t)=\bs{\mathcal T}\cdot\exp[-\bs{\mathcal D}t]\cdot
\bs{\mathcal T}^{-1}=\exp[-\bs\omega t]$ where
$\bs\omega=\bs{\mathcal T}\cdot\bs{\mathcal D}\cdot\bs{\mathcal T}^{-1}$.
It remains to be shown that $\bs\omega$ indeed coincides with
the Mori-Zwanzig expression, Eq.~\eqref{eq:omegas}.

To do so, we use the following relations between the expansion coefficients
$A_{2m}^{2n}$ and $B_{2m}^{2n+1}$, and the eigenvalues $a_{2n}(k)$ and
$b_{2n+1}(k)$ that follow from inserting the expansion Eq.~\eqref{eq:mathieuexpand} into the Mathieu differential equation \eqref{eq:mathieudgl},
\begin{subequations}
\begin{gather}\label{eq:mathieuABab}
  a_{2n}A_{2m}^{2n}=m^2A_{2m}^{2n}+k(1+\delta_{m,1})A_{2m-2}^{2n}+kA_{2m+2}^{2n}\,,\\
  b_{2n+1}B_{2m}^{2n+1}=(m+1)^2B_{2m}^{2n+1}+kB_{2m-2}^{2n+1}+kB_{2m+2}^{2n+1}\,,
\end{gather}
\end{subequations}
with the convention that $A_{2m}^{2n}=B_{2m}^{2n+1}=0$ for $m<0$. Then, a
straightforward calculation reveals that indeed
\begin{equation}
  \sum_{l''=-\infty}^\infty\mathcal T_{ll''}\mathcal D_{l''l''}\mathcal T^{-1}_{l''l'}=\left(D_tq^2+D_rl^2\right)\delta_{ll'}-\frac{iv_0q}{2}\delta_{|l-l'|,1}\,,
\end{equation}
which confirms the equivalence between the Mathieu-function expansion
of Ref.~\cite{Kurzthaler.2018} and the present Mori-Zwanzig projection,
i.e., between the series expansion Eq.~\eqref{eq:mathieuseries} and
the matrix exponential Eq.~\eqref{eq:phimatrixexp}.
Both representations remain challenging numerically; the authors of
Ref.~\cite{Kurzthaler.2018} used a cutoff of $N=40$ in the series
equivalent to Eq.~\eqref{eq:mathieuseries}, evaluating the recurrence
relation eq.~\eqref{eq:mathieuABab} with a cutoff of $100$.
Likewise, as shown in the main text, direct evaluation of the matrix
exponential requires matrices of size $(2L+1)\times(2L+1)$ with $L\simeq10$
in order to achieve similar accuracy.

\end{appendix}

\bibliography{lit}
\bibliographystyle{apsrev4-2}

\end{document}